\documentclass[final,3p,times]{elsarticle}
\usepackage{lineno}
\usepackage{graphicx}
\usepackage{amsmath}
\usepackage{natbib}
\usepackage{subcaption}
\usepackage{xcolor}
\usepackage[normalem]{ulem}
\usepackage[shortlabels,inline]{enumitem}
\usepackage{natbib}
\usepackage{tabularx}
\newcolumntype{b}{X}
\newcolumntype{s}{>{\hsize=.5\hsize}X}
\usepackage{bm}
\usepackage{bbm}
\usepackage{multirow}

\colorlet{revision}{black}

\newcommand\Rey{\mbox{Re}}           % Reynolds number

\journal{Journal of Computational Physics}

\begin{document}

\begin{frontmatter}

\title{The volume-filtering immersed boundary method}
\author[aff1]{Himanshu Dave}
\author[aff1]{Marcus Herrmann}
\author[aff1]{M. Houssem Kasbaoui\corref{cor1}}
\cortext[cor1]{Corresponding author, email: houssem.kasbaoui@asu.edu}
\address[aff1]{
School for Engineering of Matter, Transport and Energy,
Arizona State University,
Tempe,
85281,
AZ,
USA}

\begin{abstract}
  We present a novel framework to deal with static and moving immersed boundaries (IB) based on volume-filtering. In this strategy, called Volume-Filtering Immersed Boundary (VFIB) method, transport equations are derived by filtering the Navier-Stokes equations and accounting for stresses at the solid-fluid interface. The result is that boundary conditions that normally apply on the solid-fluid interface are transformed into bodyforces that apply on the right-hand side of the filtered transport equations. In this method, the filter width acts as a parameter that controls the level of resolution. The IB is considered well-resolved if the filter width is much smaller than the characteristic corrugation scale of the interface. There are several innovations in this IB method. First, it sheds light on the role of the internal flow which arises when the transport equations are solved inside the IB. We show that, it is essential to separate stresses due to the external and internal fluids in order to get accurate forces, and provide a method to do so. Second, we show that the volumes associated with Lagrangian forcing points on the boundary depend on the local topology of the surface. We provide a straightforward way to compute these volumes using a triangle tessellation of the interface and the surface density function. Third, we provide an efficient procedure to compute the solid volume fraction, thus, enabling tagging interior/exterior cells. This volume fraction is also involved in the  procedure to separate stresses due to the external fluid from the total stresses. Fourth, we show a path forward to extend the VFIB method to Large Eddy Simulations involving IBs. Lastly, we apply the VFIB in several numerical tests involving two- and three- dimensional static and moving IBs. We show greatly improved results compared to prior IB methods. Further, we test several filter kernels and show that, for well-resolved IBs, the choice of the kernel plays little role.

\end{abstract}

\begin{keyword}
Immersed boundary method \sep Volume-filtering \sep Fully-resolved simulations \sep CFD
\end{keyword}

\end{frontmatter}

\section{Introduction}
\label{sec_intro}
Besides few canonical flows, most fluid dynamic applications involve bounding surfaces with complex topology. These surfaces may be fixed, as in the case of airfoils and ship hulls, or moving, as in the case of stirred-tank reactors, flapping wings, and turbines. Despite increasing computing power, numerical simulations of such flows hinge on the availability of methods capable of capturing flow-surface interactions with accuracy while also meeting practical computational considerations such as robustness, ease of implementation, and scalability. The seminal work of Peskin \citep{peskinFlowPatternsHeart1972,peskinFluidDynamicsHeart1982,laiImmersedBoundaryMethod2000,peskinImmersedBoundaryMethod2002}, and later extended by several investigators, shows that it is possible to impose boundary conditions on topologically complex bounding surfaces without resorting to body-fitted meshes.  This approach enables the use of fast and scalable Cartesian grid solvers. Boundary conditions are imposed using ad-hoc forcing terms added to the right-hand side of the governing equations. Conceptually, the equations to be solved for an incompressible flow are
\begin{eqnarray}
	\nabla \cdot \bm{u}&=&0\\
	\frac{\partial \bm{u}}{\partial t} + \bm{u}\cdot \nabla \bm{u} &=& - \frac{1}{\rho}\nabla p+\nu \nabla^2\bm{u} + \bm{F}_\mathrm{IB} \label{eq:peskin}
\end{eqnarray}
where $\rho$ and $\nu$ are the fluid's constant density and kinematic viscosity, respectively. The ad-hoc numerical term $\bm{F}_\mathrm{IB}$ represents the immersed boundary (IB) forcing and is crafted to impose no-slip boundary conditions.  Despite the popularity of IB methods, the ad-hocness of these methods remains a problem: the forcing term $\bm{F}_\mathrm{IB}$ does not correspond to any physical term in the original Navier-Stokes equations. This makes it difficult to confidently answer long-standing questions such as
\begin{enumerate*}[(i)]
	\item should the forcing be applied to the entire volumetric region occupied by the immersed object or be limited to the solid-fluid boundary?
	\item how is the solution affected by the sharpness of the immersed boundary representation?, and
	\item how to properly compute the hydrodynamic force on the solid when the immersed boundary is diffuse?
\end{enumerate*}
Further, the fact that the IB forcing does not derive from analytical expressions makes it difficult to identify changes that would materially improve solution quality, extend the approach to Large-Eddy Simulations, or connect with established multiphase flow models in flows laden with a multitude of small immersed solids. In this manuscript, we remove the ad-hocness by deriving a new framework for immersed boundary methods that has sound theoretical footing. We show that the immersed boundary forcing can be derived rigorously by filtering the Navier-Stokes equations, discuss its discretization, and show that several other terms may be missing in previously proposed immersed boundary methods.

The work of \citet{uhlmannImmersedBoundaryMethod2005} represents a notable improvement over Peskin's original method. \citet{uhlmannImmersedBoundaryMethod2005} builds the immersed boundary forcing on a cloud of Lagrangian markers placed on the surface of the solid. Then, using convolutions with regularized Dirac delta \citep{romaAdaptiveVersionImmersed1999}, the Lagrangian forcing is transformed into the Eulerian forcing field $\bm{F}_{\mathrm{IB}}$. The Lagrangian forcing is built such that, in principle, the interpolated fluid velocities at the Lagrangian markers match the solid's velocity at these locations. \citet{uhlmannImmersedBoundaryMethod2005} shows that the method performs well in several benchmark tests with static and moving boundaries. Spurious oscillations of the hydrodynamic force observed with other immersed boundary methods \citep{uhlmannFirstExperimentsSimulation2003,leeSourcesSpuriousForce2011,seoSharpinterfaceImmersedBoundary2011,schneidersAccurateMovingBoundary2013} are significantly reduced. This makes Uhlmann's method remarkably stable even with thousands of fully resolved moving spheres. Since then, several improvements have been proposed. \citet{yangSmoothingTechniqueDiscrete2009} introduced a smoothing technique for the discrete Dirac delta that further reduces spurious oscillations observed with moving immersed boundaries. Recognizing that the diffuse nature of the IB forcing causes the IB forcing from one Lagrangian marker to affect the calculation of IB forcing on neighboring markers, \citet{luoFullscaleSolutionsParticleladen2007} proposed a variant, called multidirect forcing, where the IB forcing is imposed iteratively to improve the convergence of the Lagrangian marker velocity towards the desired no-slip velocity. \citet{breugemSecondorderAccurateImmersed2012} used this method in simulations with resolved spheres, and noticed that the Lagrangian markers must be retracted inwards to get correct hydrodynamic forces on the immersed spheres. \citet{kempeImprovedImmersedBoundary2012} proposed a variant similar to the multidirect forcing of \citet{luoFullscaleSolutionsParticleladen2007} where the IB forcing is applied iteratively, and introduced a different approach for computing hydrodynamic forces on immersed solids based on level-set functions. \citet{kasbaouiDirectNumericalSimulations2021} proposed a semi-implicit time integration scheme for the calculation of the IB forcing term based on an iterative Crank-Nicolson scheme. This approach improves the convergence of the Lagrangian marker velocity and was shown to compare well with experimental data even for inertially stirred turbulent flows in closed vessels.

Despite the success of the aforementioned methods, there are still open questions  stemming from the ad-hocness of these methods. First, the role of the internal flow inside the immersed solid is not yet fully understood. 
This flow may develop when the IB forcing is applied only on the surface of the immersed solid, leaving internal cells unforced. 
\textcolor{revision}{With the so-called fictitious domain IB methods, a rigidity constraint is applied to enforce rigid body motion within the solid \citep{glowinskiDistributedLagrangeMultiplier1999,sharmaFastComputationTechnique2005}.} However,
\citet{uhlmannImmersedBoundaryMethod2005} reports that applying the forcing throughout the volumetric region does not change his results significantly compared to when the forcing is applied on the boundary only. Thus, the latter option is preferred due to its lower computational cost. Despite being considered an artificial byproduct of the forcing technique \citep{hoflerNavierStokesSimulationConstraint2000,uhlmannImmersedBoundaryMethod2005,kempeImprovedImmersedBoundary2012}, the internal flow is used to compute hydrodynamic forces on moving spherical particles \textcolor{revision}{\citep{uhlmannImmersedBoundaryMethod2005,kempeImprovedImmersedBoundary2012,tschisgaleNoniterativeImmersedBoundary2017}}. \citet{uhlmannImmersedBoundaryMethod2005} assimilates the flow inside the sphere to rigid body motion despite it not being the case. \citet{kempeImprovedImmersedBoundary2012} show that it is necessary to embrace the non-rigid motion inside the sphere in order to compute the hydrodynamic force accurately. If the internal flow was to be neglected or zeroed out on the basis that it is artificial, then the hydrodynamic force computed by \citet{kempeImprovedImmersedBoundary2012} and \citet{uhlmannImmersedBoundaryMethod2005} would vanish.

Another ambiguous point in this class of IB methods concerns the determination of Lagrangian marker volumes which arise in the computation of the Lagrangian immersed boundary forcing. \citet{uhlmannImmersedBoundaryMethod2005} \textcolor{revision}{ and several others (e.g. \citep{kempeImprovedImmersedBoundary2012,tschisgaleNoniterativeImmersedBoundary2017})} relate the marker volume $\Delta V_m$ to the grid spacing $\Delta x$ following $\Delta V_m\sim \Delta x^3$, and uses enough markers to form a thin shell around the immersed boundary. However, \citet{zhouSpatioTemporalResolutionImmersed2021} argue that it is a misconception to pin $\Delta V_m$ to $\Delta x^3$ and suggest that it should be considered as a relaxation factor that controls how fast errors in the no-slip boundary conditions decay. They provide an expression to compute $\Delta V_m$ based on stability analysis, but, the calculation is cumbersome and only applicable to spherical immersed boundaries. In both \citep{uhlmannImmersedBoundaryMethod2005} and \citep{zhouSpatioTemporalResolutionImmersed2021}, geometric information about the immersed boundary such as surface area and curvature are not taken into consideration. However, the fact that \citet{zhouSpatioTemporalResolutionImmersed2021} find that $\Delta V_m$ must decrease with decreasing marker spacing suggests that the marker volume depends, at least in part, on the portion of the IB surface area around the marker.

To overcome the issues raised above, we derive a new framework for immersed boundary methods that does not rely on any ad-hoc elements. The method we present, called Volume-filtering Immersed Boundary (VFIB) method, relies on the volume-filtering technique introduced by \citet{andersonFluidMechanicalDescription1967a}. The method is similar in spirit to Large Eddy Simulations, where the Navier-Stokes equations are filtered with a filter kernel having width $\delta_f$. This procedure yields filtered governing equations where the effects of boundary conditions appear as right-hand side terms involving surface integrals on the immersed boundary. The derivation of these equations is physically and mathematically rigorous, and does not depend on any numerical considerations. Traditionally, volume-filtering has been applied to derive continuum equations for multiphase flows \citep{andersonFluidMechanicalDescription1967a,andersonFluidMechanicalDescription1968,jacksonDynamicsFluidizedParticles2000} and porous media \citep{whitakerFlowPorousMedia1986}. In these applications, the width of the filter kernel, $\delta_f$, is chosen much larger than the characteristic interface corrugation scale $\delta_c$, such as particle radius or pore size. In this way,  Eulerian quantities representing solid volume fractions, mass, and momentum can be extracted from large ensembles of discrete solids. For the purpose of deriving an immersed boundary method, we take the opposite limit: the filter width is much smaller than the characteristic interface corrugation scales. In this limit ($\delta_f\ll \delta_c$), the immersed boundary as well as all flow scales larger than $\delta_f$ are fully resolved. The resulting equations form the basis of the VFIB method. As we show in this manuscript, the VFIB method generalizes Uhlmann's immersed boundary method and sheds much needed clarification on the role of the internal flow, Lagrangian marker volume, and a more accurate way to compute hydrodynamic forces.

\textcolor{revision}{
The objectives of the present manuscript are two-fold. The first goal is to present the theory underpinning the Volume-filtering Immersed Boundary method. This is done in section \ref{sec:derivation}, where we present the derivation of the governing equations that are solved in the VFIB method. The emphasis here is on the mathematical and physical framework which does not depend on the choice of numerical parameters.  With the theory clearly established, the second goal of this manuscript is to provide an implementation of the VFIB method, discuss practical considerations (such as the choice of filter kernel, relative size of filter kernel and mesh spacing, calculation of volume fraction, and forces on the IB), and demonstrate the approach in canonical test cases. We address the discretization and numerical implementation of these equations in section \ref{sec:implementation}. In section \ref{sec:filter_analysis}, we discusses the characteristics of different filter kernels tested in the present study. We illustrate the approach using numerical examples with static, forcibly, and freely moving immersed boundaries in section \ref{sec:test_cases}. Finally, we give concluding remarks in section \ref{sec:conclusion}.}

\section{The Volume-Filtering Immersed Boundary Method}
\label{sec:derivation}

In this section, we apply the volume-filtering method of \citet{andersonFluidMechanicalDescription1967a} to the problem of a fluid with an immersed solid. We consider two formulations and discuss the merits of both. In the \emph{two-phase} formulation, we describe the dynamics of the flow outside the immersed boundary separately from the dynamics inside the solid. In the \emph{one-phase} formulation, we consider that the immersed object is hollow and filled with the same fluid as outside of it. By describing the total mass and momentum conservation, we arrive at governing equations that can be discretized efficiently and where the Immersed Boundary forcing is given explicitly.

\subsection{Two-phase formulation}
\begin{figure}
	\centering
	\includegraphics[width=0.9\linewidth]{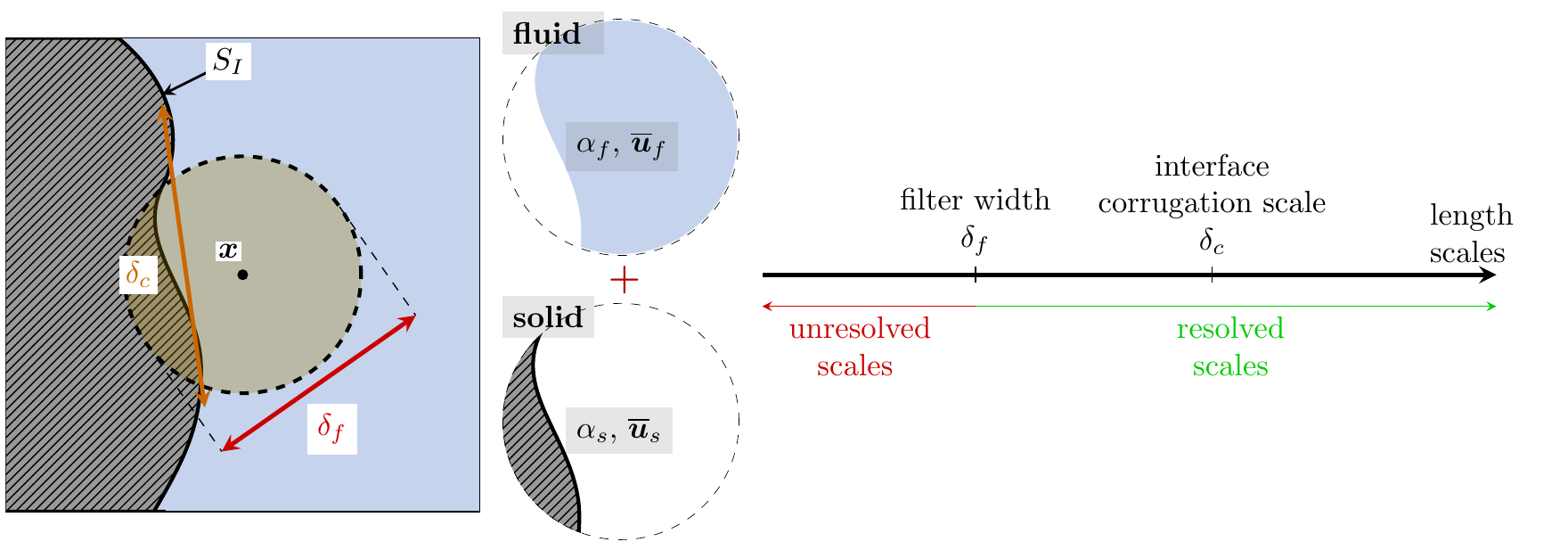}
	\caption{Illustration of the volume-filtering approach. Filtering the point-wise fields allows the extraction of average fluid and solid volume fractions ($\alpha_f$ and $\alpha_s$), mass ($\alpha_f\rho_f$ and $\alpha_s\rho_s$) and momentum ($\alpha_f\rho_f\overline{\bm{u}}_f$ and $\alpha_s\rho_s\overline{\bm{u}}_s$) within a region of size $\delta_f$.  The immersed boundary is well resolved when the characteristic corrugation scale $\delta_c$ of the interface is much larger than the filter width $\delta_f$.}
	\label{fig:schematic_VF}
\end{figure}
Consider two-phases separated by an interface $S_I$ as shown in the schematic in figure \ref{fig:schematic_VF}. Phase ``$f$'' corresponds to the fluid phase, while phase ``$s$'' corresponds to the immersed solid. Where the fluid \emph{exists}, mass and momentum conservation are given by the incompressible Navier-Stokes equations
\begin{align}
  \nabla\cdot\bm{u}_f&\;=\;0\label{eq:RT1_1}\\
  \rho_f\left(\frac{\partial\bm{u}_f}{\partial t}+\nabla\cdot(\bm{u}_f\bm{u}_f)\right)&\;=\;\nabla\cdot\bm{\tau}_f\label{eq:RT1_2}
\end{align}
where $\bm{u}_f$, $p$, and $\bm{\tau}_f=-p\bm{I}+\mu_f (\nabla \bm{u}_f+\nabla \bm{u}^T_f)$ are the fluid velocity, pressure, and stress tensor, respectively. Throughout, the manuscript we assume that the fluid density $\rho_f$ and viscosity $\mu_f$ are constant. Borrowing the terminology from \citet{jacksonDynamicsFluidizedParticles2000} equations (\ref{eq:RT1_1}) and (\ref{eq:RT1_2}) are called the \emph{point-wise} conservation equations.

For the familiar Navier-Stokes equations (\ref{eq:RT1_1}) and (\ref{eq:RT1_2}), the boundary conditions are imposed through additional constraints that apply on the fluid-solid interface $S_I$ and outside boundaries.
Considering an impermeable interface, no slip-boundary conditions apply, i.e.,
\begin{align}
 \bm{u}_f(\bm{x})&\;=\;\bm{u}_{I}(\bm{x})\qquad\text{for}\;\bm{x}\in S_I\label{eq:RT1_3}
\end{align}
where $\bm{u}_{I}(\bm{x})$ is the velocity of a point $\bm{x}$ located on the interface.

To illustrate how volume-filtering works, consider a point $\bm{x}$ near the immersed boundary as schematized in figure \ref{fig:schematic_VF}. By filtering the point-wise quantities, we can extract the fluid volume fraction $\alpha_f$, average fluid mass $\alpha_f\rho_f$, and fluid momentum $\alpha_f\rho_f\overline{\bm{u}}_f$ in a region of size $\delta_f$ around the point $\bm{x}$. In this way, we can define the filtered fluid velocity $\overline{\bm{u}}_f$ at location $\bm{x}$ corresponding to the average fluid velocity under the filter kernel. As shown in figure \ref{fig:schematic_VF}, it is only the volumetric region occupied by the fluid that counts towards the extracted fluid quantities. Likewise, we can extract the solid volume fraction $\alpha_s$, average solid mass $\alpha_s\rho_s$, and solid momentum $\alpha_s\rho_s\overline{\bm{u}}_s$ around the point $\bm{x}$ by considering only the region occupied by the solid that falls under the filter kernel. However, because we do not seek to model the internal solid dynamics, we will focus on the fluid phase only. Contrary to the point-wise velocity $\bm{u}_f$, which exists only in the fluid region, the filtered fluid velocity $\overline{\bm{u}}_f$ can be defined for any arbitrary point in space. Moving the probing point $\bm{x}$  towards the solid causes the filtered fluid velocity $\overline{\bm{u}}_f$ to decay smoothly to zero as the fluid region under the filter kernel shrinks. Note that the filtered fluid velocity at points $\bm{x}$ inside the solid, but less than $\delta_f/2$ away from $S_I$, may not be zero as there is still a fluid region under the filter kernel. Thus, volume-filtering smears the effect of the interface over a region of size $\delta_f$. The immersed boundary is well resolved when the filter width $\delta_f$ is much smaller than the characteristic corrugation scale $\delta_c$ of the immersed boundary. For example, if the immersed solid is a sphere of diameter $D$, we would require that $\delta_f\ll \delta_c=D$ to retain good resolution of the immersed boundary. Length scales below $\delta_f$ are unresolved and must be modeled. In the limit where $\delta_f$ is vanishingly small, the filtered velocity becomes a discontinuous field equal to the point-wise velocity inside the fluid and zero inside the solid.

To formalize this idea, we consider a \emph{symmetric} filter kernel $g$ that integrates to \emph{unity} and has \emph{compact} support of width $\delta_f$. Mathematically, these properties read,
\begin{align}
    \iiint_{\bm{y}\in \mathbbm{R}^3} g(\bm{y})dV&\;=\;1,&\text{(unitary)}\label{eq:filter_def}\\
    g(-\bm{y})&\;=\;g(\bm{y}),&\text{(symmetric)}\\
    g(\bm{y})&\;=\;0\ \mbox{if}\ ||\bm{y}||\geq \delta_f/2.&\text{(compact)}
\end{align}
Notice that the integration is considered over the entire space which includes regions occupied by both fluid and solid phases. Symmetry of the kernel is required to avoid artificial anisotropy and plays an important role in the derivation of the filtered conservation equations. Compactness of the filter kernel serves primarily a computational purpose as it allows fast numerical integration of $g$ on surfaces, but is otherwise not required for the purpose of the derivation.

The fluid volume fraction at any arbitrary location $\bm{x}$ is given by
\begin{align}
  \alpha_f(\bm{x},t)&\;=\;\iiint_{\bm{y}\in \mathbbm{R}^3}\mathbbm{1}_{f}(\bm{y},t)g(\bm{x}-\bm{y})dV\label{eq:RT1_4},
\end{align}
where $\mathbbm{1}_f(\bm{y},t)$ is an indicator function equal to 1 if $\bm{y}$ is in the fluid and 0  otherwise. Similarly, the solid volume fraction $\alpha_s$ is defined by replacing the fluid indicator function in equation (\ref{eq:RT1_4}), with that of the solid, i.e., $\mathbbm{1}_s=1-\mathbbm{1}_f$. The volume fraction $\alpha_f(\bm{x})$ represents the ratio of the volume occupied by the fluid to the total volume under the support of the filter kernel centered at a location $\bm{x}$. Regions of space occupied exclusively by the fluid have a fluid volume fraction $\alpha_f=1$. Conversely, regions where only the solid phase can be found within the support of the filter kernel have a fluid volume fraction $\alpha_f=0$. Locations where $0<\alpha_f(\bm{x})<1$ correspond to those where both phases are within reach of the filter kernel, as in the schematic in figure \ref{fig:schematic_VF}. This corresponds to a narrow band of width $\delta_f$ around the interface. Isocontours $\alpha_f=\alpha_s= 0.5$ give the location of the interface, if the latter can be considered locally planar. In this formulation, discontinuous effects across the interface are smoothed over a distance $\delta_f$. The interface representation can be made sharper by reducing the size of $\delta_f$. Equation (\ref{eq:filter_def}) guarantees that the solid and fluid volume fractions add up to unity at any given location, i.e., $\alpha_f(\bm{x})+\alpha_s(\bm{x})=1$ for any $\bm{x}\in\mathbbm{R}^3$. At this point, we emphasize that the volume fraction in equation (\ref{eq:RT1_4}), filtered velocities, and filtered governing equations to follow are not tied to any specific discretization or mesh.

 For an arbitrary point $\bm{x}$ in space, the volume-filtered fluid velocity $\overline{\bm{u}}_f$ is defined as following,
\begin{eqnarray}
  \alpha_f(\bm{x},t)\overline{\bm{u}}_f(\bm{x},t)&=&{ \iiint_{\bm{y}\in\mathbbm{R}^3} \mathbbm{1}_f(\bm{y},t)\bm{u}_f(\bm{y},t)g(\bm{x}-\bm{y})dV}.
  \label{eq:filtered_u}
\end{eqnarray}
 The volume-filtered velocity is \emph{continuous} and \emph{exists everywhere}. It tends smoothly to 0 a distance $\delta_f/2$ away within the solid.

 The volume-filtered governing equations are derived by application of the filter to the point-wise equations (\ref{eq:RT1_1}) and (\ref{eq:RT1_2}). Using the divergence theorem and symmetry of the filter kernel ($\partial g(\bm{x}-\bm{y})/\partial x_i= -\partial g(\bm{x}-\bm{y})/\partial y_i$), one can show that filtering the gradient, divergence and time derivative operators yields,
\begin{eqnarray}
  \alpha_f(\bm{x}) \overline{\nabla \bm{\Lambda}}(\bm{x}) &=&\nabla (\alpha_f \overline{\bm{\Lambda}}) - \iint_{\bm{y}\in S_I} \bm{n}\bm{\Lambda}(\bm{y},t) g(\bm{x}-\bm{y})dS,\label{eq:identity1}\\
  \alpha_f(\bm{x}) \overline{\nabla\cdot \bm{\Lambda}}(\bm{x}) &=&\nabla \cdot (\alpha_f \overline{\bm{\Lambda}}) - \iint_{\bm{y}\in S_I} \bm{n}\cdot\bm{\Lambda}(\bm{y},t) g(\bm{x}-\bm{y})dS,\label{eq:identity2}\\
  \alpha_f(\bm{x}) \overline{\frac{\partial \bm{\Lambda}}{\partial t}}(\bm{x}) &=&\frac{\partial (\alpha_f \overline{\bm{\Lambda}})}{\partial t} + \iint_{\bm{y}\in S_I} (\bm{n}\cdot\bm{u})\bm{\Lambda}(\bm{y},t) g(\bm{x}-\bm{y})dS,\label{eq:identity3}
\end{eqnarray}
where $\bm{\Lambda}$ is an arbitrary vector or tensor property of the fluid. Here, $\bm{n}$ represents the normal vector at the interface pointing from the solid to the fluid phase. A full derivation of these identities can be found in the original work of \citet{andersonFluidMechanicalDescription1967a}. Because the application of the filter removes the notion of a boundary ($\bm{x}$ can be anywhere in $\mathbbm{R}^3$), information from the boundary conditions emerges in identities (\ref{eq:identity1}), (\ref{eq:identity2}), and (\ref{eq:identity3}) as surface integrals on the interface separating the solid and fluid phases.

Applying the filtering procedure to the mass and momentum equations (\ref{eq:RT1_1}) and (\ref{eq:RT1_2}) leads to
\begin{eqnarray}
  \frac{\partial\alpha_f}{\partial t}+\nabla\cdot(\alpha_f\overline{\bm{u}_f})&=&0,\label{eq:RT1_5}\\
  \rho_f\left(\frac{\partial}{\partial t}(\alpha_f\overline{\bm{u}_f})+\nabla\cdot(\alpha_f\overline{\bm{u}_f\bm{u}_f})\right)&=&\nabla\cdot(\alpha_f\overline{\bm{\tau}}_f)-\iint_{\bm{y}\in S_{I}} \bm{n}\cdot \bm{\tau}_f(\bm{y},t) g(\bm{x-y})dS.\label{eq:RT1_6}
\end{eqnarray}
In the filtered momentum equation (\ref{eq:RT1_6}), the term
\begin{equation}
	\bm{F}_{I,f}(\bm{x},t)=\iint_{\bm{y}\in S_{I}} \bm{n}\cdot \bm{\tau}_f(\bm{y},t) g(\bm{x-y})dS
\end{equation}
represents a force density exerted by the immersed solid on the fluid. Note that this term includes a surface integral on the immersed boundary, meaning that the forcing is limited to a thin region of width $\delta_f$ around the immersed boundary rather than the entire volumetric region. The force exerted by the fluid on the solid can be obtained by integrating the immersed boundary force density over the entire domain (fluid and solid regions):
\begin{equation}
	\iint_{\bm{y}\in S_{I}} \bm{n}\cdot \bm{\tau}_f(\bm{y},t)dS= \iiint_{\bm{x}\in\mathbbm{R}^3}\bm{F}_{I,f}(\bm{x},t)dV.
\end{equation}
The relationship above does not require $\delta_f=0$, and holds true for arbitrary filter widths and surface curvatures. The filtered stress tensor $\alpha_f\overline{\bm{\tau}_f}=\alpha_f[\overline{-p\bm{I}+\mu_f (\nabla\bm{u}_f+\nabla\bm{u}_f^T)}]$ requires detailed examination. Whereas filtering the pressure part is straightforward ($-\alpha_f\overline{p \bm{I}}= -\alpha_f \overline{p}\bm{I}$), filtering the viscous part leads to
\begin{eqnarray}
  \mu_f \alpha_f\left(\overline{\nabla \bm{u}_f}+\overline{\nabla \bm{u}_f}^T\right)&=&
  \mu_f \alpha_f\left(\nabla \overline{ \bm{u}}_f+\nabla \overline{\bm{u}}_f^T\textcolor{revision}{-\frac{2}{3}(\nabla\cdot\overline{\bm{u}}_f)\bm{I}}\right)+\alpha_f\bm{R}_{\mu,f}
  \label{eq:viscous_stresses}
\end{eqnarray}
where $\bm{R}_{\mu,f}$ represents the residual viscous stress tensor.
% and reads
%{\color{revision}
%\begin{equation}
%	\bm{R}_{\mu,f}(\bm{x})=\mu_f\left\{\iint_{\bm{y}\in S_{I}} \Bigl(
%     \bm{n}(\bm{y})(\overline{\bm{u}}_f(\bm{x})-\bm{u}_f(\bm{y}))
%     +
%     (\overline{\bm{u}}_f(\bm{x})-\bm{u}_f(\bm{y}))\bm{n}(\bm{y})
%     \Bigl)
%     g(\bm{x-y})dS
%     +\alpha_f \frac{2}{3}\nabla\cdot\overline{\bm{u}}_f\right\}.
%\end{equation}
%}
Equation (\ref{eq:viscous_stresses}) is derived by application of the identity (\ref{eq:identity1}). The residual viscous stress tensor $\mathbf{R}_{\mu,f}$ is zero away from the immersed boundary. Near the IB, the effects of $\mathbf{R}_{\mu,f}$ may be significant if the IB is poorly resolved. In the limit where $\delta_f\rightarrow 0$, the residual viscous stresses vanish.

 Application of the filter leads to the emergence of unclosed convective terms $\alpha_f\overline{\bm{u}_f\bm{u}_f}$ in equation (\ref{eq:RT1_6}). To deal with these terms, we introduce the subfilter-scale stress tensor,
\begin{equation}
 \bm{\tau}_{\mathrm{sfs},f} =\overline{\bm{u}_{f}\bm{u}_{f}} - \overline{\bm{u}}_{f}\overline{\bm{u}}_{f}.
 \label{eq:tau_sfs}
\end{equation}

To summarize, the set of equations obtained by volume-filtering equations (\ref{eq:RT1_1}) and (\ref{eq:RT1_2}) are:
\begin{eqnarray}
  \frac{\partial\alpha_f}{\partial t}+\nabla\cdot(\alpha_f\overline{\bm{u}_f})&=&0, \label{eq:twophase_1}\\
  \rho_f\left(\frac{\partial}{\partial t}(\alpha_f\overline{\bm{u}}_f)+\nabla\cdot(\alpha_f\overline{\bm{u}}_f\,\overline{\bm{u}}_f)\right)&=&\nabla\cdot\left(-\alpha_f\overline{p}\bm{I}+\alpha_f\mu_f\left(\nabla\overline{\bm{u}}_f+\nabla\overline{\bm{u}}_f^T\textcolor{revision}{-\frac{2}{3}(\nabla\cdot\overline{\bm{u}}_f)\bm{I}}\right)+\alpha_f\bm{R}_{\mu,f}\right)
    	-\bm{F}_{I,f}-\nabla \cdot (\alpha_f\bm{\tau}_\mathrm{sfs,f}).
  	\hspace{1cm}
  	\label{eq:twophase_2}
\end{eqnarray}
While the derivation of  equations (\ref{eq:twophase_1}) and (\ref{eq:twophase_2}) is mathematically rigorous, discretizing and solving these equations in their present form is challenging. The main complication is due to the fluid volume fraction that vanishes inside the solid. Because of this, the filtered fluid velocity $\overline{\bm{u}}_f$ cannot be computed from the transported quantity $(\alpha_f\overline{\bm{u}}_f)$ without leading to large errors and computational instabilities.

\textcolor{revision}{Lastly, as we discuss in \ref{sec:multiphase_form}, equations (\ref{eq:twophase_1}) and (\ref{eq:twophase_2}) can be further transformed to arrive at a form that is frequently used in multiphase flows, see equation (\ref{eq:appendix_3}). However, such procedure is not needed for the present purpose}.

\subsection{One-phase formulation}
\label{sec:thin_boundary}

We now present an alternative formulation that overcomes the numerical stability issues encountered with the two-phase formulation. In the one-phase formulation, we assume that the immersed solid is hollow and filled with fluid having identical density and viscosity as the fluid outside. In this view, the immersed boundary represents an infinitely thin interface that separates two fluids. We denote $\alpha_1$ and $\alpha_2$ the volume fractions occupied by the exterior and interior fluids, respectively. Applying volume-filtering to both fluids, we obtain the following mass and momentum conservation equations,
\begin{eqnarray}
  \frac{\partial\alpha_i}{\partial t}+\nabla\cdot(\alpha_i\overline{\bm{u}}_i)&=&0,\label{eq:vf_1}\\
  \rho_f\left(\frac{\partial(\alpha_i\overline{\bm{u}}_i)}{\partial t}+\nabla\cdot(\alpha_i\overline{\bm{u}}_i\,\overline{\bm{u}}_i)\right)&=&\nabla\cdot\left(-\alpha_i\overline{p}_i\bm{I}+\alpha_i\mu_f\left(\nabla\overline{\bm{u}}_i+\nabla\overline{\bm{u}}_i^T\textcolor{revision}{-\frac{2}{3}(\nabla\cdot\overline{\bm{u}}_i)\bm{I}}\right)+\alpha_i\bm{R}_{\mu,i}\right)-\bm{F}_{I,i}-\nabla \cdot(\alpha_i \bm{\tau}_{\mathrm{sfs},i}).\label{eq:vf_2}
\end{eqnarray}
where $i = 1$ or 2 depending on which fluid is considered. In addition to the above equations, the volume fractions are constrained by the condition $\alpha_1 + \alpha_2 = 1$.

To obtain single-field equations, we sum over the two fluids in equations (\ref{eq:vf_1}) and (\ref{eq:vf_2}):
\begin{eqnarray}
\nabla\cdot\bm{u}_m&=&0,\label{eq:vf_3}\\
  \rho_f\left(\frac{\partial\bm{u}_m}{\partial t}+\nabla\cdot(\bm{u}_m\,\bm{u}_m)\right)&=&-\nabla p_m+\mu_f\nabla^2\bm{u}_m-\bm{F}_{I,m}+\nabla \cdot (\bm{\tau}_{r}-\bm{\tau}_{\mathrm{sfs,m}}).\label{eq:vf_4}
\end{eqnarray}
These equations describe the transport of the total mass and momentum of both fluids. Here, $\bm{u}_m=\alpha_1 \overline{\bm{u}}_1+\alpha_2\overline{\bm{u}}_2$ is the mixture velocity, $p_m=\alpha_1 \overline{p}_1+\alpha_2\overline{p}_2$ is the mixture pressure, $\bm{F}_{I,m}=\bm{F}_{I,1}+\bm{F}_{I,2}$, and $\bm{\tau}_{\mathrm{sfs},m}=\bm{\tau}_{\mathrm{sfs},1}+\bm{\tau}_{\mathrm{sfs},2}$. The viscous stress tensors and residual stress tenors $\bm{R}_{\mu,i}$ combine to give the term $\mu_f\nabla^2\bm{u}_m$ in equation (\ref{eq:vf_4}). For ease of notation, we drop the subscript $m$ in the rest of the manuscript.

In this one-phase formulation, we see the emergence of a new tensor,
\begin{equation}
  \bm{\tau}_r= \bm{u}\bm{u}-\sum_i \alpha_i\overline{\bm{u}}_i\overline{\bm{u}}_i=\alpha_1\alpha_2( \overline{\bm{u}}_1-\overline{\bm{u}}_2)(\overline{\bm{u}}_2-\overline{\bm{u}}_1).
\end{equation}
The tensor $\bm{\tau}_r$ represents the momentum drift across the interface. This tensor can be neglected for sufficiently well resolved immersed boundaries, i.e., $\delta_f/\delta_c\ll 1$. Away from the interface, $\bm{\tau}_r=0$  since $\alpha_1=0$ or $\alpha_2=0$. Near the interface, no-slip boundary conditions lead to $\overline{\bm{u}}_1\simeq\overline{\bm{u}}_2\simeq\bm{u}_I$, where $\bm{u}_I$ is the interface velocity, hence, $\bm{\tau}_r\simeq 0$. In the rest of the paper, we will assume that the immersed boundary representation is sufficiently sharp such that $\bm{\tau}_r$ can be considered identically zero everywhere. This assumption is equivalent to stating $\alpha_1\alpha_2(\overline{\bm{u}}_1-\overline{\bm{u}}_2)=0$, and allows us to extract the internal and external fluid velocities using:
\begin{eqnarray}
  \alpha_1\overline{\bm{u}}_1&=& \alpha_1\bm{u} + \alpha_1\alpha_2(\overline{\bm{u}}_1-\overline{\bm{u}}_2)\simeq\alpha_1 \bm{u}\\
  \alpha_2\overline{\bm{u}}_2&=& \alpha_2\bm{u} + \alpha_1\alpha_2(\overline{\bm{u}}_2-\overline{\bm{u}}_1)\simeq\alpha_2 \bm{u}
\end{eqnarray}

In the one-phase formulation, the subfilter scale term is
\begin{equation}
 \bm{\tau}_{\mathrm{sfs}} = \sum_i \alpha_i\bm{\tau}_{\mathrm{sfs},i} = \alpha_{1}(\overline{\bm{u}_{1}\bm{u}_{1}}-\overline{\bm{u}}_{1}\overline{\bm{u}}_{1})+ \alpha_{2}(\overline{\bm{u}_{2}\bm{u}_{2}}-\overline{\bm{u}}_{2}\overline{\bm{u}}_{2})
\end{equation}
In general, the subfilter-scale tensor must be closed if the immersed boundary is poorly resolved ($\delta_f/\delta_c=O(1)$ or $\delta_f/\delta_c\gg 1$). In particle-laden flows, where similar volume-filtering is carried out using filter kernels typically much larger than the particle diameters, the subfilter scale stresses are known as pseudo-turbulent stresses and are subject of active research and modeling \citep{senRolePseudoturbulentStresses2018,mooreLagrangianInvestigationPseudoturbulence2019}. Closures may also be required if the flow scales $\delta_u$ are not well-resolved ($\delta_f/\delta_u\gtrsim 1$), even if the immersed boundary is well-resolved $\delta_f/\delta_c\ll 1$. For example, if two well-resolved immersed boundaries approach one another, but the gap between the two solids becomes smaller than the filter width $\delta_f$, $\bm{\tau}_\mathrm{sfs}$ should be augmented with a lubrication model \citep{costaCollisionModelFully2015} to represent the effect of the subfilter fluid in the gap between the two objects. In the case where the flow is turbulent and the filter width is larger than the turbulence scales,  closure of $\bm{\tau}_\mathrm{sfs}$, away from the immersed boundary, may be carried out by any of the well-established LES models such as turbulent eddy viscosity model of \citet{boussinesqTheorieAnalytiqueChaleur1901}, the dynamic Smagorinsky model \citep{germanoDynamicSubgridScale1991,lillyProposedModificationGermano1992}, or scale-similarity models \citep{bardinoImprovedTurbulenceModels1983}. While there is an array of closure models available to model $\bm{\tau}_\mathrm{sfs}$, this paper focuses on the validation of the VFIB method in the limit of well resolved immersed boundaries ($\delta_f/\delta_c\ll 1$) and flow scales ($\delta_f/\delta_u\ll 1$) , and therefore the subfilter-scale terms will be neglected.

The total immersed boundary force density can be written as
\begin{eqnarray}
  \bm{F}_{I}&=& \bm{F}_{I,1}+\bm{F}_{I,2} =
   \iint_{S_I} \left(\bm{n}\cdot\left[\bm{\tau}_1-\bm{\tau}_2\right]\right)(\bm{y})g(\bm{x}-\bm{y})dS \label{eq:172}
\end{eqnarray}
where $\bm{n}$ is the normal pointing from the internal fluid ($i=2$) to the external fluid ($i=1$). It is important to acknowledge that this term accounts for forces exerted on the immersed boundary by both fluids inside and outside. Computing the hydrodynamic force on an immersed object requires extracting the contribution of the external fluid, $\bm{F}_{I,1}$, from the total immersed boundary force density $\bm{F}_{I}$. This aspect is addressed in \S \ref{sec:appendix_force}.

The force density $\bm{F}_{I}$ can be expressed explicitly using mixture quantities. To do so, consider the Taylor series of the total surface stresses in the vicinity of a point $\bm{y}_I$ located at the interface
\begin{eqnarray}
  \left(\bm{n}\cdot\left[\bm{\tau}_1-\bm{\tau}_2\right]\right)(\bm{y}) &=&
  	\left(\bm{n}\cdot\left[\bm{\tau}_1-\bm{\tau}_2\right]\right)(\bm{y}_I)
  	+(\bm{y}-\bm{y}_I)\cdot \nabla \left(\bm{n}\cdot\left[\bm{\tau}_1-\bm{\tau}_2\right]\right)(\bm{y}_I) +O(\delta_f^2)
	\label{eq:taylor_series}
\end{eqnarray}
Introducing equation (\ref{eq:taylor_series}) in equation (\ref{eq:vf_4}) and rearranging the terms leads to
\begin{eqnarray}
  \left(\bm{n}\cdot\left[\bm{\tau}_1-\bm{\tau}_2\right]\right)(\bm{y}_I) =
  	- \ell(\bm{y}_I)\left.\left( \rho_f\frac{D\bm{u}}{Dt} - \nabla\cdot\left(-p\bm{I}+\mu_f (\nabla u+\nabla u^T)+\bm{\tau}_r-\bm{\tau}_{\mathrm{sfs}}\right) \right)\right|_{\bm{y}=\bm{y}_I} +O(\delta_f^2)
  	\label{eq:interface_stress}
\end{eqnarray}
where $D/Dt$ is the \textcolor{revision}{substantial derivative}. Thanks to the symmetry of the filter kernel $g$, the first order term from the Taylor series cancels out leadings to a second order approximation in $\delta_f$ of the immersed boundary stresses.

In equation (\ref{eq:interface_stress}), the length
\begin{eqnarray}
  \ell(\bm{y}_I)= \left(\iint_{S_I} g(\bm{y}_I-\bm{y})dS\right)^{-1} \label{eq:smearing}
\end{eqnarray}
 represents the interface smearing length at a location $\bm{y}_I$ on the immersed boundary. The smearing length is on the same order as the filter kernel width $\delta_f$, however, the exact value depends on both interface curvature and filter kernel $g$. If the interface around point $\bm{y}$ can be considered flat, then $\ell(\bm{y})=g(\bm{0})^{-1/3}$. For the triangle and cosine filter kernels described in \S \ref{sec:filter_analysis}, the smearing length for a flat interface is $\ell(\bm{y})=2/\delta_f$ and $\pi/(2\delta_f)$, respectively.

The smearing length is closely connected to the interface surface density. For a point $\bm{y}$ on the interface, equation (\ref{eq:smearing}) shows that $\ell(\bm{y})$ is the inverse of the surface density $\Sigma$ at that location, where
\begin{eqnarray}
  \Sigma(\bm{x})= \iint_{S_I} g(\bm{x}-\bm{y})dS.\label{eq:SDF}
\end{eqnarray}
The surface density $\Sigma$ represents how much surface area of the immersed boundary is under the filter kernel. Away from the interface, $\Sigma=0$. Near the interface, the value of $\Sigma$ depends on the choice of filter kernel, width $\delta_f$, and local curvature. Integrating $\Sigma$ over the entire domain gives the total surface area of the immersed boundary $A_\mathrm{I}$:
\begin{equation}
	\iiint_{\bm{x}\in\mathbbm{R}^3}\Sigma(\bm{x})dV=A_\mathrm{I}.
\end{equation}
\textcolor{revision}{Note that reducing $\delta_f$ increases $\Sigma$ and, consequently, reduces the smearing length $\ell$ computed at a point on the interface.}

{\color{revision}
At this point, the no-slip boundary condition can be introduced in equation (\ref{eq:interface_stress}). To do so, notice that a fluid particle located on the interface has an acceleration that matches the acceleration of the IB,
\begin{equation}
	\frac{D}{Dt}\bm{u}(\bm{y}) = \frac{d}{dt}\bm{u}_I(\bm{y})\quad \bm{y}\in S_I. \label{eq:vfib_bc}
\end{equation}

Once expressions (\ref{eq:interface_stress}) and (\ref{eq:vfib_bc}) are introduced in (\ref{eq:172}), we see that the interface stress is given by:
\begin{eqnarray}
  \bm{F}_{I}(\bm{x})= -\iint_{S_I} \ell(\bm{y})\left.\left( \rho_f\frac{d\bm{u}_I}{dt} + \nabla p -\mu_f\nabla^2 \bm{u} -\nabla\cdot\left(\bm{\tau}_r-\bm{\tau}_{\mathrm{sfs}}\right) \right)\right|_{\bm{y}}g(\bm{x}-\bm{y})dS +O(\delta_f^2) \label{eq:201}
\end{eqnarray}
Thus, the complete equations that must be solved in the one-phase VFIB method are
\begin{eqnarray}
\nabla\cdot\bm{u}&=&0,\label{eq:vfib_1}\\
  \rho_f\left(\frac{\partial\bm{u}}{\partial t}+\nabla\cdot(\bm{u}\,\bm{u})\right)&=&-\nabla p+\mu_f\nabla^2\bm{u}+\nabla \cdot (\bm{\tau}_{r}-\bm{\tau}_{\mathrm{sfs}}) 
  + \iint_{S_I} \ell\left( \rho_f\frac{d\bm{u}_I}{dt}  + \nabla p -\mu_f\nabla^2 \bm{u} -\nabla\cdot\left(\bm{\tau}_r-\bm{\tau}_{\mathrm{sfs}}\right)\right)gdS.\nonumber\\[-1ex]\label{eq:vfib_2}
\end{eqnarray}

The most significant difference between the one-phase VFIB method and the IB method introduced by Peskin, is that the present method intentionally accounts for the effect of smearing the IB. The thickness of the smearing is controlled using the filter thickness $\delta_f$. Notice that, in the limit of vanishing filter thickness ($\delta_f\rightarrow 0$), we recover Peskin's equations as the unclosed terms in equation (\ref{eq:vfib_2}) vanish and the filter kernel $g$ becomes a Dirac delta ($g\rightarrow\delta$). However, there are many advantages to choosing non-zero filter thickness. From a modeling perspective, this enables us to account for the effect of under-resolved flow motion in the vicinity of the IB using sub-filter scale terms $\bm{\tau_r}$ and $\bm{\tau}_\mathrm{sfs}$, if models can be supplied. This also allows us to purposely coarsen the resolution in applications such as LES, by taking $\delta_f$ larger than some cut-off length scale of the flow. From a computational perspective, the choice of $\delta_f\neq0$ removes the need to discretize and regularize a Dirac delta distribution. This also has the advantage of decoupling the choice of filter width $\delta_f$ from the choice of grid spacing $\Delta x$, or any other discretization parameter. It also offers several advantages in the calculation of stresses on the IB and volume fraction field as we shall see in section \ref{sec:implementation}.
}

\subsection{Volume fraction computation}\label{sec:vol_frac}
Although the volume fractions $\alpha_1$ and $\alpha_2$ do not appear explicitly in the one-phase formulation (equations (\ref{eq:vf_1}) and (\ref{eq:vf_2})), the computation of the volume fraction serves three goals:
\begin{enumerate*}[(i)]
	\item distinguish between interior and exterior points,
	\item compute the total volume occupied by the immersed solid, and most importantly
	\item extract the respective contributions due to external and internal fluids $\bm{F}_{I,1}$ and $\bm{F}_{I,2}$ from the total immersed boundary force density $\bm{F}_{I}$ (see \S \ref{sec:appendix_force}).
\end{enumerate*}

Rather than using the formal definition equation (\ref{eq:RT1_4}) which requires computationally expensive procedures to build an indicator function \citep{apteNumericalMethodFully2009,kempeImprovedImmersedBoundary2012}, we compute the external fluid volume fraction $\alpha_1$ by solving a Poisson equation. To derive this equation, first, replace $\bm{\Lambda}$ with the identity tensor $\bm{I}$ in equation (\ref{eq:identity2}) to obtain the gradient of the external fluid volume fraction,
\begin{eqnarray}
  \nabla \alpha_1(\bm{x},t)&=& \iint_{\bm{y}\in S_I} \bm{n} g(\bm{x}-\bm{y})dS.
\end{eqnarray}
The Poisson equation for the volume fraction is obtained by taking the divergence of the equation above:
\begin{eqnarray}
\color{revision}  \nabla^2 \alpha_1(\bm{x},t)&=& \nabla\cdot\iint_{\bm{y}\in S_I} \bm{n} g(\bm{x}-\bm{y})dS. \label{eq:vf_poisson}
\end{eqnarray}
The above Poisson equation derived by volume-filtering is similar to the Poisson equation used to build phase-indicator functions in simulations of bubbly flows with the front-tracking method of \citet{unverdiFronttrackingMethodViscous1992}. Equation (\ref{eq:vf_poisson}) can be parallelized and solved efficiently using elliptic solvers, including in simulations where immersed boundaries move and require frequent updates of $\alpha_1$. \textcolor{revision}{In the examples provided in section \ref{sec:test_cases}, we use an algebraic multigrid method to solve equation (\ref{eq:vf_poisson}) with Dirichlet boundary conditions on the domain boundaries. For simplicity, we solve for the volume fraction throughout the entire domain, although it would suffice to solve (\ref{eq:vf_poisson}) in a narrow band of thickness $\delta_f$ around the solid-fluid interface similar to the approach of \citet{unverdiFronttrackingMethodViscous1992}. Note that in the case of static immersed boundaries, equation (\ref{eq:vf_poisson}) needs to be solved only once. For forcibly moved IBs, the solver (equations (\ref{eq:vfib_1}) and (\ref{eq:vfib_2})) does not require the volume fraction, and as such, the computation of the volume-fraction is carried out as part of the post-processing workflow. For cases of Fluid-Structure Interaction, such as freely-moving particles, the volume fraction is needed at each step in order to compute the force on the immersed solids and update their dynamics (see example in \S\ref{sssec:freely_falling})}.
The volume fraction of the internal region is found using $\alpha_2=1-\alpha_1$.

{\color{revision}
\subsection{Computing the force due to the external fluid}
\label{sec:appendix_force}

As discussed in the introduction, the role of the internal flow must be clarified. In the one-phase formulation, the internal flow is not an artificial byproduct of the forcing technique, but instead has physical meaning. In the present approach, we explicitly consider hollow solids filled with the same fluid as the one they have been immersed into. As a result, if the solid moves, the internal fluid develops a non-zero velocity, causing additional stresses on the interface. Even if the solid is static, the average nature of the formulation makes it that solving the mixture equations (\ref{eq:vf_3}) and (\ref{eq:vf_4}) may lead to non-zero mixture velocities inside the hollow solid, especially if the interface is very diffuse (i.e., $\delta_f/\delta_c$ is not $\ll 1$). Thus, it is essential to be able to isolate the hydrodynamic stresses due to the external fluid from the total hydrodynamic stresses due to both internal and external fluids.

 If stresses due to the internal flow can be neglected, the force on the immersed solid would be
\begin{eqnarray}
	\hspace{-0.5cm}\iint_{S_I}\bm{n}\cdot\bm{\tau}_1dS\hspace{-0.2cm}&=&\hspace{-0.2cm}\iiint_{\mathbbm{R}^3} \bm{F}_{I,1}dV\\
	&\simeq &\hspace{-0.2cm} \iiint_{\mathbbm{R}^3}\bm{F}_I dV=  -\iint_{\bm{y}\in S_I} \ell(\bm{y})\left.\left( \rho_f\frac{D}{D t}\bm{u} + \nabla p -\mu_f\nabla^2\bm{u}  -\nabla\cdot(\bm{\tau}_r-\bm{\tau}_\mathrm{sfs})\right)\right|_{\bm{y}}dS
\end{eqnarray}
This approximation may be acceptable for flows with static boundaries and high momentum, since the internal flow is generally comparatively very small.

A more accurate estimation of the external forces on the immersed solid can be obtained by isolating and calculating the stresses due to the external fluid directly. Starting from the momentum conservation equations (\ref{eq:vf_2}) for the external flow ($i=1$) and following the same approach as in section \ref{sec:thin_boundary}, we obtain the force due to the external flow,
\begin{eqnarray}
  \iint_{S_I}\bm{n}\cdot\bm{\tau}_1dS= -\iint_{\bm{y}\in S_I} \left.\ell(\bm{y})\left( \rho_f \alpha_1\frac{D}{D t}\overline{\bm{u}}_1 - \nabla\cdot\left(-\alpha_1 \overline{p}_1\bm{I} +\mu_f \alpha_1\left(\nabla \overline{\bm{u}}_1+\nabla \overline{\bm{u}}_1^T\textcolor{revision}{-\frac{2}{3}(\nabla\cdot\overline{\bm{u}}_1)\bm{I}}
  \right)+ \alpha_1\bm{R}_{\mu,1}  -\alpha_1\bm{\tau}_{\mathrm{sfs},1}\right) \vphantom{\frac{D\overline{\bm{u}}_1}{D t}} \right)\right|_{\bm{y}}dS \label{eq:37}
\end{eqnarray}
This expression can be rearranged in the following way
\begin{eqnarray}
  \hspace{-0.7cm}\iint_{S_I}\bm{n}\cdot\bm{\tau}_1 dS &=&
  \iint_{\bm{y}\in S_I}
  \bm{n}\cdot(\overline{\bm{\tau}}_1-\bm{\tau}_{\mathrm{sfs},1}) dS  -\iint_{\bm{y}\in S_I} \alpha_1(\bm{y})\ell(\bm{y})\left.\left(
	\rho_f \frac{D}{D t}\overline{\bm{u}}_1 -  \nabla\cdot\left(\overline{\bm{\tau}}_1-\bm{\tau}_{\mathrm{sfs},1}\right)
  \right)\right|_{\bm{y}} dS \label{eq:47}
\end{eqnarray}
where $\overline{\bm{\tau}}_1=-\overline{p}_1\bm{I} +\mu_f (\nabla \overline{\bm{u}}_1+\nabla \overline{\bm{u}}_1^T-2/3\nabla\cdot\overline{\bm{u}}_1)+\bm{R}_{\mu,1}$ is the filtered stress tensor.

Since we solve for the mixture quantities ($\bm{u}$ and $p$), it is more advantageous to express (\ref{eq:47}) in terms of these quantities. This can be done when the immersed boundary is well-resolved. Under such condition, we may neglect the tensors $\bm{\tau}_\mathrm{sfs}$ and $\bm{R}_\mu$, and make the approximations $\overline{\bm{u}}_1(\bm{y})\simeq\bm{u}(\bm{y})$ and $\overline{p}_1(\bm{y})\simeq p(\bm{y})$ for points $\bm{y}$ on the interface $S_I$. Thus, equation (\ref{eq:47}) becomes
\begin{eqnarray}
	\iint_{S_I}\bm{n}\cdot\bm{\tau}_1dS&\simeq&
	\iint_{S_I} \left( -p\bm{I} +\mu_f (\nabla \bm{u}+\nabla \bm{u}^T) \right)\cdot \bm{n}dS
	-\iint_{\bm{y}\in S_I} \alpha_1(\bm{y}) \ell(\bm{y})\left.\left( \rho_f\frac{d}{d t}\bm{u}_I + \nabla p-\mu_f\nabla^2\bm{u} \right)\right|_{\bm{y}} dS, \label{eq:53}
\end{eqnarray}
where we have also used the no-slip boundary condition in the form $D\bm{u}/Dt=d\bm{u}_I/dt$.

Equation (\ref{eq:53}) shows that, to get the force exerted on the immersed boundary, it is not enough to integrate the (resolved) stresses on the immersed boundary, i.e, computing the first term on the right-hand side of (\ref{eq:53}) only. Doing so would lead to significant under-estimation of the force due to the external fluid. This is, perhaps, the reason why prior investigators sought alternative ways to compute the stresses on the immersed boundary, for example, by converting $\iint_{S_I}\bm{n}\cdot\bm{\tau}_1dS$ into a volumetric integral and involving the internal flow. Such methods are not needed if both terms in (\ref{eq:53}) are computed. This is the method that we use in the rest of the manuscript to calculate hydrodynamic forces on immersed boundaries.
}
%. It is this form that we use in the rest of the manuscript to calculate hydrodynamic forces on immersed boundaries.

%In our tests, we find that the contribution of the residual viscous stress to the subfilter scale stresses (second integral in the right-hand side of (\ref{eq:53})) is negligible. Thus, this integral can be directly evaluated using the immersed boundary forcing at centroids determined in step 3 (see section \ref{sec:time_integration}), considerably simplifying the calculation. In summary, the force on the immersed solid at the $n+1$ time step is:
%\begin{eqnarray}
%	\iint_{S_I}\bm{n}\cdot\bm{\tau}_1^{n+1}dS&\simeq &
%	 \sum_{m=1}^N \left\{\left(-p^{n+1} +\mu_f ( \nabla\bm{u}^{n+1}+ (\nabla\bm{u}^{n+1})^T)
%		+\frac{\bm{R}_{\mu}^{n+1}}{\alpha^{n+1}}\right)\cdot\bm{n}_m A_m\right\}\nonumber\\
%	&-&\sum_{m=1}^N\left\{ \alpha^{n+1}_1(\bm{x}_m^{n+1})\ell(\bm{x}^{n+1}_m)\left(\frac{\bm{u}^{n+1}_{I,m}-\widetilde{\bm{u}}^{n+1}_{k+1}(\bm{x}_m^{n+1})}{\Delta t}\right)A_m\right\}.
%\end{eqnarray}

\section{Numerical implementation}\label{sec:implementation}
The VFIB method is implemented in a library called LEAP and interfaced with the flow solver NGA. Below, we describe only elements pertaining to the implementation of the VFIB method. Details about other aspects of the flow solver NGA can be found in \citep{desjardinsHighOrderConservative2008}.
\subsection{Spatial discretization of the interface}
\begin{figure}
	\centering
	\includegraphics[width=0.5\linewidth]{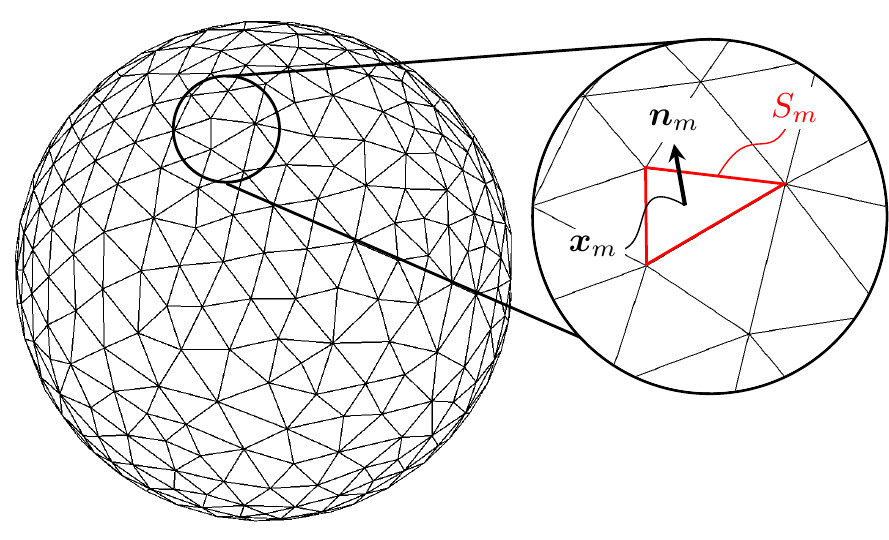}
	\caption{Example of a mesh of a spherical immersed boundary. The ``$m$''-th mesh triangle, $S_m$, has surface area $A_m$, centroid $\bm{x}_m$, and outward pointing normal $\bm{n}_m$.
	\label{fig:sphere_mesh}}
\end{figure}
  We now focus on the discretization of the forcing term in equation (\ref{eq:201}).
Suppose that the immersed interface has been meshed, such that $S_I=\cup_{m=1}^N S_m$, where $S_m$ are the elements of the mesh. Figure \ref{fig:sphere_mesh} shows an example of a mesh for a spherical immersed boundary. The IB forcing can be written as the sum of discrete contributions from each mesh element
  \begin{equation}
    \bm{F}_{I}(\bm{x})= -\sum_{m=1}^N \iint_{\bm{y} \in S_m} \ell(\bm{y})\left.\left( \rho_f\frac{d}{dt}\bm{u}_I + \nabla p -\mu_f\nabla^2
    \bm{u} -\nabla\cdot\left(\bm{\tau}_r-\bm{\tau}_{\mathrm{sfs}}\right) \right)\right|_{\bm{y}}g(\bm{x}-\bm{y})dS+O(\delta_f^2).
  \end{equation}
  Assuming that the typical mesh width is $O(\Delta x)$ and using the mid-point rule leads to
  \begin{eqnarray}
  &&\bm{F}_{I}(\bm{x})= -\sum_{m=1}^N\left\{ \ell(\bm{x}_m)\left.\left( \rho_f\frac{d}{dt}\bm{u}_I + \nabla p -\mu_f\nabla^2
    \bm{u} -\nabla\cdot\left(\bm{\tau}_r-\bm{\tau}_{\mathrm{sfs}}\right) \right)\right|_{\bm{x}_m}g(\bm{x}-\bm{x}_m)A_m \right\} \nonumber \\
  &&\hspace{9.5cm}+ O(\Delta x^2,\delta_f^2) \label{eq:ib_forcing}.
  \end{eqnarray}
  where $\bm{x}_m$ is the location of the centroid of the mesh element $S_m$ and $A_m$ its surface area.

Expression (\ref{eq:ib_forcing}) suggests a possible interpretation where the triangle centroids are viewed as Lagrangian forcing points, similar to the view adopted by \citet{uhlmannImmersedBoundaryMethod2005}. With this interpretation, each Lagrangian forcing point in (\ref{eq:ib_forcing}) can be associated with a volume $\Delta V_m=\ell(\bm{x}_m)A_m$. Thus, in the VFIB method, the Lagrangian point volume depends on the local curvature of the immersed boundary, choice of filter kernel $g$, and centroid spacing through the smearing length $\ell$ and the triangle surface area $A_m$. This is in contrast with the method of \citet{uhlmannImmersedBoundaryMethod2005} where Lagrangian points are assigned a fixed volume  $\Delta V_m=\Delta x^3$, regardless of the local topology of the immersed boundary. \textcolor{revision}{Further, the volume $\Delta V_m=\ell(\bm{x}_m)A_m$ can be computed directly from the surface density ($\Sigma=\ell^{-1}$) and the triangle surface area $A_m$ without resorting to any optimization method as done by \citet{zhouSpatioTemporalResolutionImmersed2021}.}

\subsection{Temporal discretization}\label{sec:time_integration}
The time integration scheme  is based on a semi-implicit iterative Crank-Nicolson scheme originally developed by \citet{akselvollLargeeddySimulationTurbulent1996} and  \citet{pierceProgressvariableApproachLargeeddy2004a} and recently adapted by \citet{kasbaouiDirectNumericalSimulations2021} for simulations with immersed boundaries. The steps below describe the update from time $t^n$ to $t^{n+1}$.

  \textbf{Step 1a:}
  The first step consists in updating the immersed boundary to the $n+1$ time step. This is performed by updating the locations and velocities of the Lagrangian centroids to the new time step.
  \begin{eqnarray}
    \bm{x}_m^n\rightarrow  \bm{x}_m^{n+1};\quad
    \bm{u}_{I,m}^n\rightarrow \bm{u}_{I,m}^{n+1}
  \end{eqnarray}
  In cases where the motion of the boundary is predetermined, the positions and velocities can be updated according to the laws of rigid body motion. A more general scheme can be used for applications in fluid-structure interaction.

  \textbf{Step 1b:}
  Once the new location of the centroids is found, we compute the surface density $\Sigma$ using:
	\begin{equation}
		\Sigma^{n+1}(\bm{x})=\sum_m^N g(\bm{x}-\bm{x}_m^{n+1})A_m
	\end{equation}
  which represents equation (\ref{eq:SDF}) discretized with the mid-point rule. Section \ref{sec:appendix_interp} provides details on the extrapolation procedure used to build the fields $g(\bm{x}-\bm{x}_m)$ in our Finite Volume solver.

  \textbf{Step 1c:}
  Next, the smearing length at the centroids is obtained by taking the inverse of the surface density interpolated at the centroids,
	$$\ell(\bm{x}^{n+1}_m)=\left(\Sigma^{n+1}(\bm{x}_m^{n+1})\right)^{-1}.$$
  Interpolations are performed by taking convolutions with the filter kernel $g$. Details are provided in \S \ref{sec:appendix_interp}.

  \textbf{Step 1d:}
  If desired, the new fluid volume fraction field $\alpha_1^{n+1}$ is computed by solving the discretized Poisson equation (\ref{eq:vf_poisson}),
  {\color{revision}
  \begin{equation}
    \nabla^2 \alpha_1^{n+1}= \nabla \cdot \left\{\sum_m^N  \bm{n}_m g(\bm{x}-\bm{x}_m)A_m\right\}.
  \end{equation}
   }

 Note that if the immersed boundary is static, steps 1a-d need only be performed at the simulation start.

  \textbf{Step 2:} At this step the iterative loop is initiated.  We assume that $k$ sub-iterations have been performed and show the calculations for $(k+1)^\mathrm{th}$ sub-iteration. As in \citep{kasbaouiDirectNumericalSimulations2021}, operator splitting is used to decouple momentum update, immersed boundary forcing, and pressure correction. At this step, the momentum update is performed without the immersed boundary force density:
  \begin{eqnarray}
    \bm{u}_{k}^{n+1/2} &=& \left(\bm{u}_{k}^{n+1}+\bm{u}^{n}\right)/2\\
    \widetilde{\bm{u}}^{n+1}_{k+1}&=& \bm{u}^n +\Delta t \left(
     -\left.\nabla\cdot(\bm{u}\bm{u})\right|^{n+1/2}_k-\frac{1}{\rho_f}p^{n+1}_k
     + \frac{\mu_f}{\rho_f}\nabla^2\bm{u}^{n+1/2}_{k+1}
     + \frac{\partial \mathcal{M}}{\partial \bm{u}}\left(\frac{ \widetilde{\bm{u}}^{n+1}_{k+1}-\bm{u}^{n+1}_{k} }{2}\right)
     \right).\label{eq:mom_update}
  \end{eqnarray}
  In the above, the operator $\mathcal{M}$ represents the sum of the convective and viscous operators,
  \begin{equation}
    \mathcal{M}(\bm{u})=-\nabla\cdot(\bm{u}\bm{u})+\frac{\mu_f}{\rho_f}\nabla^2\bm{u}.
  \end{equation}
  The Jacobian $\partial M/\partial \bm{u}$ in equation (\ref{eq:mom_update}) allows the treatment of the non-linearity with a Newton-Raphson method \citep{pierceProgressvariableApproachLargeeddy2004a}. The momentum equation is solved using the approximate factorization technique of \citet{choiEffectsComputationalTime1994} based on the Alternating Direction Implicit (ADI) method.

   \textbf{Step 3:} Next, the immersed boundary term is applied. Using expression (\ref{eq:ib_forcing}), the force density $\bm{F}_I$ is discretized as 
  \begin{equation}
    \bm{F}^{n+1}_{I,k+1}(\bm{x})= \sum_{m=1}^N \left\{ \ell(\bm{x}^{n+1}_m)\left(\frac{\bm{u}^{n+1}_{I,m}-\widetilde{\bm{u}}^{n+1}_{k+1}(\bm{x}_m^{n+1})}{\Delta t}\right)g(\bm{x}-\bm{x}^{n+1}_m) A_m\right\}.
  \end{equation}
  where we have used the fact that the velocity at $n+1$ at the centroid location must match the interface velocity at the new time step $\bm{u}^{n+1}_{I,m}$ to fulfill the no-slip boundary condition. The velocity field is then updated using,
  \begin{equation}
    \widehat{\bm{u}}^{n+1}_{k+1}= \widetilde{\bm{u}}^{n+1}_{k+1} +\Delta t \bm{F}^{n+1}_{I,k+1}. \label{eq:IB_step1}
  \end{equation}

   \textbf{Step 4:} The pressure-Poisson equation is solved and a final velocity correction is applied to yield a divergence-free field,
  \begin{eqnarray}
    \nabla^2 p^{n+1}_{k+1}&=& -\frac{\rho_f}{\Delta t} \nabla\cdot\widehat{\bm{u}}^{n+1}_{k+1},\\
    \bm{u}_{k+1}^{n+1}&=&\widehat{\bm{u}}^{n+1}_{k+1}-\frac{\Delta t}{\rho_f}\nabla (p^{n+1}_{k+1}-p^{n+1}_{k}).
  \end{eqnarray}

\textbf{Step 5:} Repeat steps 2 to 4 until completion of the iterative Crank-Nicolson loop. Typically, two to three sub-iterations per time step are used \citep{kasbaouiDirectNumericalSimulations2021}.

\subsection{Interpolations and extrapolations}
\label{sec:appendix_interp}

\begin{figure}\centering
	\includegraphics[width=4in]{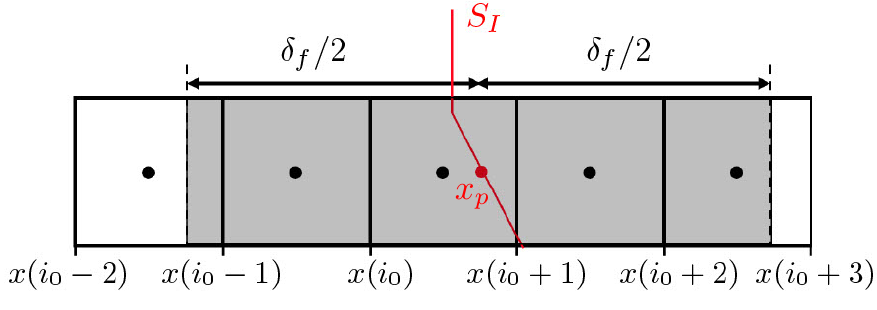}
	\caption{Schematic showing the cells involved in the 1-D interpolations and extrapolations at a point $x_p$ on the interface. In this example, the filter kernel has width $\delta_f=4\Delta x$. The shaded areas shows the cells that are under the reach of the filter.}
	\label{fig:interp_extrap}
\end{figure}

In order to implement the VFIB method, there are two operations that require close examination: (a) \emph{interpolations} of Eulerian quantities (e.g. fluid velocities) onto Lagrangian points (e.g. centroids of surface triangles), and (b) \emph{extrapolations} of Lagrangian quantities (e.g. forcing at the centroids of surface triangles) onto the grid.
\subsubsection{Interpolations}
Interpolations are carried out by taking convolutions with the filter kernel $g$. For the sake of brevity, we illustrate how the calculation is performed in a one-dimension. Figure \ref{fig:interp_extrap} shows a sketch of the configuration.

For a Lagrangian point located at $x_p$, the interpolation of an Eulerian quantity $\Lambda(x)$ at the Lagrangian point is
\begin{eqnarray}
  \Lambda(x_p)=\int_{-\infty}^{+\infty}\Lambda(x')g(x'-x_{p})dx'.\label{eq:interp_1}
\end{eqnarray}
Taking into account the compactness of $g$, i.e.,  $g(x')=0$ if $|x'-x_p| \geq \delta_f/2$, the interpolated quantity may be written as
\begin{eqnarray}
  \Lambda(x_p)=\int_{x_p -\delta_f/2}^{x_p +\delta_f/2}\Lambda(x')g(x'-x_p)dx'.\label{eq:interp_2}
\end{eqnarray}
With a filter kernel $\delta_f/\Delta x=4$, the integral above can be split into 5 contributions coming from cell $i_0$ containing the Lagrangian point and two neighboring cells on each side (see figure \ref{fig:interp_extrap}),
\begin{eqnarray}
  \Lambda(x_p)&=&\sum_{i=i_0-2}^{i_0+2}\int_{x(i)}^{x(i+1)}\Lambda(x')g(x'-x_p)dx'.\label{eq:interp_3}
\end{eqnarray}
Finally, we use a mid-point rule to get a second order approximation of the interpolated value
\begin{eqnarray}
  \Lambda(x_p)  &=& \frac{\sum_{i=i_0-2}^{i_0+2}\Lambda(x(i+1/2))g(x(i+1/2)-x_p)\Delta x(i)}
                {\sum_{i=i_0-2}^{i_0+2}g(x(i+1/2)-x_p)\Delta x(i)} \label{eq:interp_4}
\end{eqnarray}
where $\Delta x(i)= x(i+1)-x(i)$ is the size of cell $i$. Normalization in (\ref{eq:interp_4}) is needed to ensure that the kernel is unitary in a discrete sense.
\subsubsection{Extrapolations}
Extrapolating a Lagrangian quantity $\lambda_p$ defined at the Lagrangian point $\bm{x}_p$ consists in building the Eulerian field $\Lambda_p(\bm{x})=\lambda_p g(\bm{x}-\bm{x}_p)$ on the fluid grid. For illustration, we will use the same one-dimensional configuration shown in figure \ref{fig:interp_extrap}.

We perform extrapolating operations in a Finite-Volume sense, i.e., we compute the cell averages
\begin{equation}
  \Lambda_p(i) = \frac{1}{\Delta x(i)}\int_{x(i)}^{x(i+1)} \Lambda_p(x')dx'= \lambda_p\left( \frac{1}{\Delta x(i)}\int_{x(i)}^{x(i+1)} g(x'-x_p)dx'\right)
\end{equation}
for $i=i_0-2,\dots,i_0+2$. For all other $i$, $\Lambda_p(i)=0$ due to the compactness of the filter kernel. In order to reduce spurious force oscillations with moving immersed boundaries, we calculate the integrals above analytically. To do so, we use the following decomposition,
\begin{eqnarray}
  \int_{x(i)}^{x(i+1)} g(x'-x_p)dx' &=&  \int_{0}^{x_r} g(x')dx' - \int_{0}^{x_l} g(x')dx',\label{eq:extrap_1}
\end{eqnarray}
where $x_r=\min(\max(x(i+1)-x_p,-\delta_f/2),\delta_f/2)$ and $x_l=\min(\max(x(i)-x_p,-\delta_f/2)$.
With the filter kernels considered in this study, the integrals in (\ref{eq:extrap_1}) can be computed by hand with ease and directly implemented into the solver.

\section{Characteristics of different filter kernels}
\label{sec:filter_analysis}

\begin{table}\centering
\begin{tabular}{lp{10cm}}
	Kernel    & Expression \\ \hline\hline
%	Box       & $	g_1(r)= \frac{1}{\delta_f}\text{ if }|r\leq \delta_f/2|,\; 0 \text{ otherwise}$. \\[8pt]
	Triangle  & $	g_1(r)= \frac{2}{\delta_f}\left(1-2|r|/\delta_f\right)\text{ if }|r|\leq \delta_f/2,\; 0 \text{ otherwise}$. \\[8pt]
	Parabolic & $	g_1(r)= \frac{3/2}{\delta_f}\left(1-(2r/\delta_f)^2\right)\text{ if }|r|\leq \delta_f/2,\; 0 \text{ otherwise}$. \\[8pt]
	Cosine    & $	g_1(r)= \frac{\pi/2}{\delta_f}\cos(\pi r/\delta_f)\text{ if }|r|\leq \delta_f/2,\; 0 \text{ otherwise}$. \\[8pt]
	Triweight & $	g_1(r)= \frac{35/16}{\delta_f}\left(1-(2r/\delta_f)^2\right)^3\text{ if }|r|\leq \delta_f/2,\; 0 \text{ otherwise}$. \\[8pt]
	\hline
\end{tabular}
\caption{One-dimensional filter kernels considered in the present study.\label{tab:kernels}}
\end{table}

\begin{figure}\centering
	\includegraphics[width=0.7\linewidth]{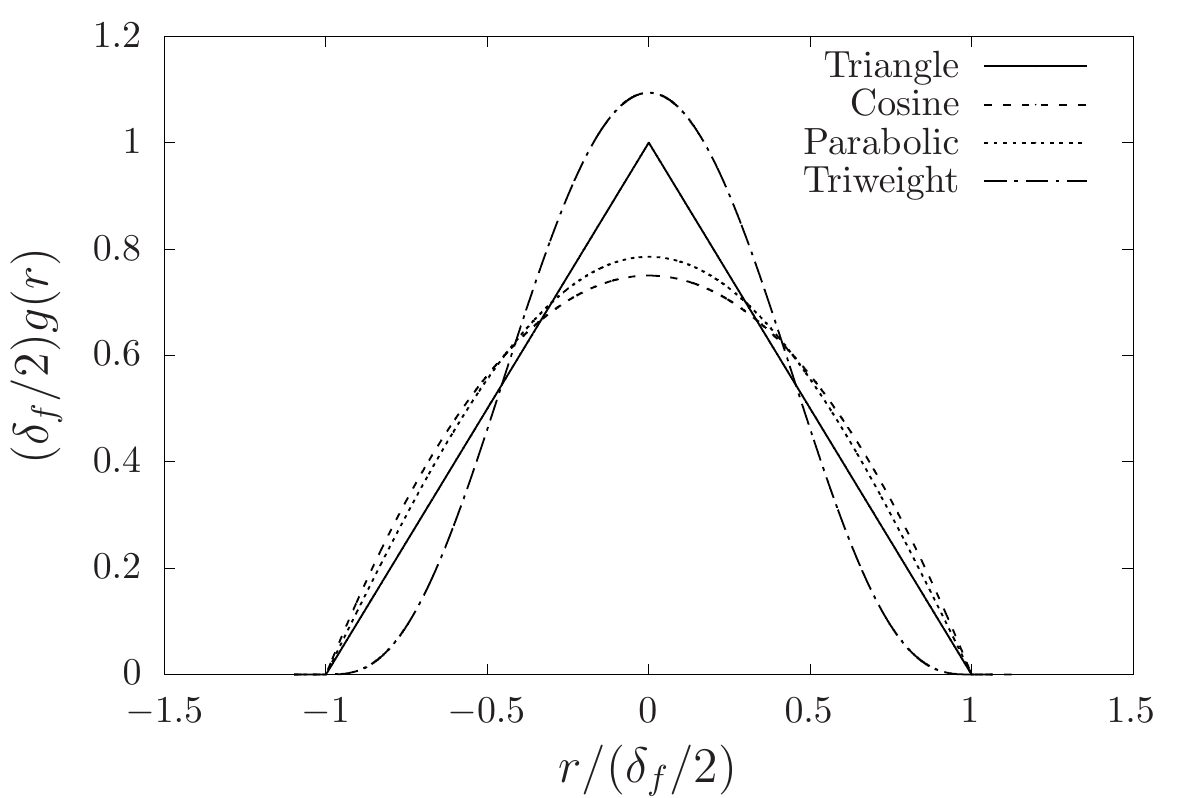}
	\caption{Graph of the one-dimensional filter kernels in table \ref{tab:kernels}. The filter kernels are symmetric, integrate to unity, and have a compact support with width $\delta_f$.}
	\label{fig:filter_type}
\end{figure}

The filter kernel plays a direct role in the immersed boundary force density, the surface density $\Sigma$, and the fluid/solid volume fraction. In the present work, we build the three-dimensional filter kernel $g$ as the  product of three 1-dimensional filter kernels
\begin{equation}
	g(\bm{x}-\bm{y})= g_1(|x_1-y_1|)g_1(|x_2-y_2|)g_1(|x_3-y_3|).
\end{equation}
We consider 4 different 1-dimensional filter kernels called triangle, parabolic, cosine, and triweight as described in table \ref{tab:kernels}.  Figure \ref{fig:filter_type} shows a graph of these kernels. These filters are unitary, symmetric, and compact with width $\delta_f$. The triweight filter has the highest peak and therefore the narrowest kernel with the majority of the weight distributed in the center, followed by the triangle kernel. Lastly, the cosine and parabolic kernel are similar in shape and have the smallest peak with the widest weight distribution. 

\textcolor{revision}{
Here, we emphasize that there is a fundamental difference between Peskin's regularized Dirac delta and the filter kernels considered here. The properties introduced by \citet{peskinImmersedBoundaryMethod2002} originate from the fact that a true distribution cannot be represented on a discrete grid, thus,  additional properties are needed to \emph{regularize} the discrete Dirac delta. In contrast, the filter kernels discussed here, must satisfy only two conditions, being unitary and symmetric, to make the derivation of the volume-filtered equations possible. The third condition of compactness is added for computational efficiency, but other filters with infinite support (e.g.  Gaussian filter) can be used in principle. With this distinction emphasized, we can also show that a unitary, symmetric, and compact filter kernel satisfies Peskin's properties in a discrete sense. Considering a 1D kernel, these three conditions yield
\begin{eqnarray}
 \text{compact:} && \quad g_1(x)=0 \quad \text{for}\quad |x|\geq \delta_f/2\\
 \text{unitary:} && \int_{\mathbbm{R}} g_1 dx =1 \rightarrow  \sum_i g_1(x_i)\Delta x_i = 1 \label{eq:filter_prop_1}\\
 \text{symmetric:} && g_1(-x)=g_1(x)\rightarrow \int_{\mathbbm{R}} (x'-x)g_1(x'-x) dx =0
  \rightarrow  \sum_i (x_i-x_{i_0}) g_1(x_i-x_{i_0}) \Delta x_i = 0 \label{eq:filter_prop_2}
\end{eqnarray}
where the integrals are discretized using the mid-point rule. 
Thus, the requirement that $g$ is unitary, symmetric, and compact is sufficient to satisfy Peskin's conditions \citep{peskinImmersedBoundaryMethod2002}. 
%Moreover, the conditions (\ref{eq:filter_prop_1}) and (\ref{eq:filter_prop_2}) above further generalize Peskin's properties \citep{peskinImmersedBoundaryMethod2002} to non-uniform grids, although we have not tested such cases yet.
}

\begin{figure}
    \centering
    \includegraphics[width=0.9\linewidth]{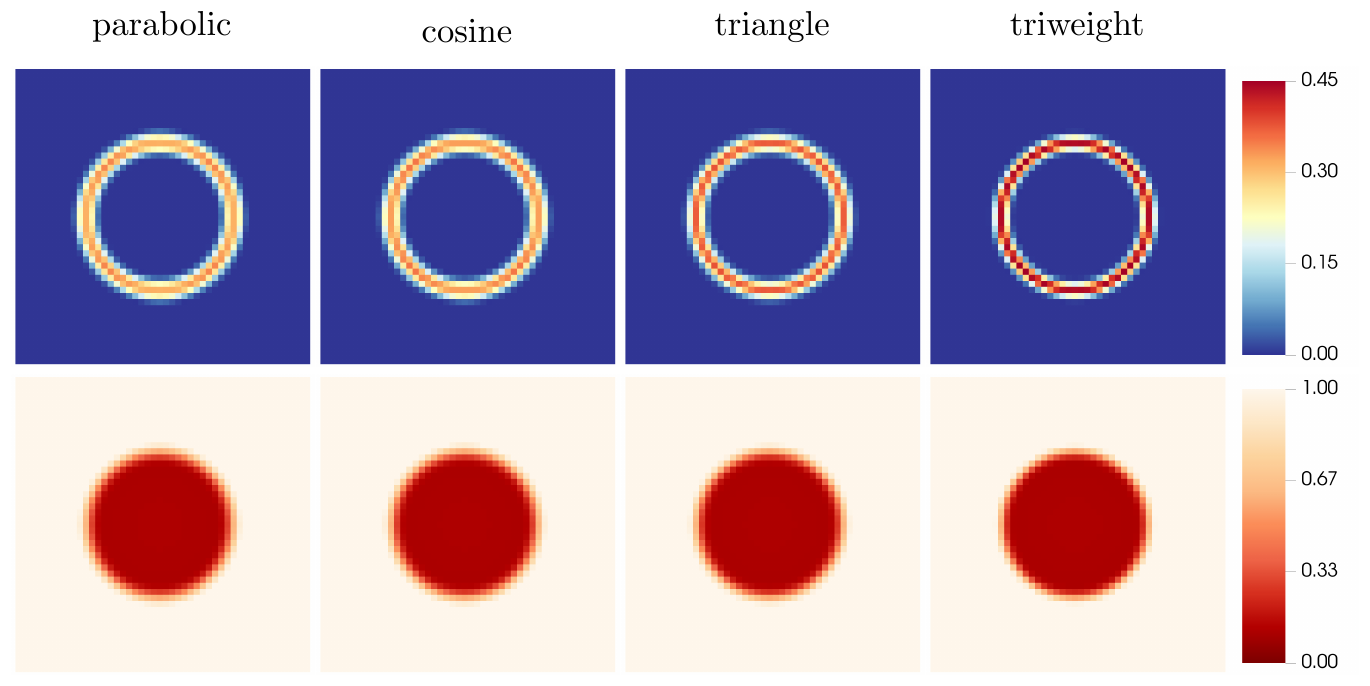}
	\caption{The effect of varying filter kernels on the (top) normalized surface density $\Sigma/(\pi D Lz/\delta_f^3)$ and (bottom) fluid volume fraction $\alpha_1$ for a circular immersed boundary of diameter $D$ with filter width $\delta_f/D = 1/6$. The grid spacing is such that 4 grid points lie across the filter width ($\delta_f/\Delta x=4$). \label{fig:filter_example}}
 \end{figure}
 To illustrate the impact of these kernels, we consider a circular immersed boundary of diameter $D$ in figure \ref{fig:filter_example}. The Cartesian grid is uniform with grid spacing $\Delta x=D/24$. In order to retain small stencils for interpolations and extrapolations, the ratio of filter width to mesh spacing is set to a fixed value $\delta_f/\Delta x = 4$. This gives a moderate resolution of the cylinder as shown by the ratio $\delta_f/D=1/6$. The top panels in figure \ref{fig:filter_example} show the surface density $\Sigma$ obtained using the four filter kernels in table \ref{tab:kernels}. The surface density field reflects clearly the relative spread of the filter kernel, whereby the narrowest kernel, the triweight kernel here, leads to the largest surface density peaks. Conversely, the parabolic kernel, which is the most spread out, leads to comparatively lower surface density. The bottom panels in figure \ref{fig:filter_example} shows the fluid volume fraction field obtained by solving the Poisson equation (\ref{eq:vf_poisson}). As expected, inside the fluid volume fraction is equal to 0 inside the cylinder and equal to 1 outside. Near the boundary, the volume fraction transitions smoothly. This transition is slightly sharper for the narrower filter kernels, although these differences are not as significant as those seen for surface density.

\section{Test cases using the VFIB method}
\label{sec:test_cases}
In this section, we apply the VFIB method in five benchmark tests with static and moving immersed boundaries. \textcolor{revision}{
The rationale for the choice of these cases is as follows:
\begin{itemize}
    \item \textbf{Fixed cylinder in a channel flow}: This two-dimensional case demonstrates the VFIB in a simple configuration. We use this case to investigate the effect of varying (i) filter kernels, (ii) varying filter sizes $D/\delta_f$, and (iii) resolutions ($D/\Delta x$ and $\delta_f/\Delta x$).
    \item \textbf{Fixed cylinder in free stream}: We consider this second two-dimensional test case to enable comparison with other immersed boundary methods since this is among the most popular benchmark tests.
    \item \textbf{Laterally oscillating cylinder in crossflow}: This case demonstrates the ability of the VFIB method to handle forcibly moving IBs in 2D.
%    \item \textbf{Flow past a NACA airfoil}: This case shows how arbitrarily shaped IBs, i.e, other than circles and spheres, can be handled using the VFIB method.
    \item \textbf{Flow past a sphere}: This case shows the ability of the method to handle 3D static IBs, and reproduce intricate wake patterns.
    \item \textbf{Freely falling sphere under gravity}: This case shows the ability of the method to handle freely moving 3D IBs. It shows the potential of the method in applications related to Particle-Resolved DNS and Fluid-Structure Interaction (FSI).
\end{itemize}
}

\subsection{Fixed cylinder in a channel flow}
\label{sssec:channel}

In this first test, we consider the two-dimensional case of a static cylinder of diameter $D$ in a channel at Reynolds number $\Rey_D=100$. The channel length and height are $L_x=22D$ and $L_y=4.1D$, respectively. The cylinder is placed asymmetrically at $x=y=0.3D$. A parabolic inflow with average velocity $U$ is prescribed at the inlet $x=0$. The average velocity $U$ and fluid kinematic viscosity $\nu$ are chosen such that $\Rey_D=UD/\nu=100$.

In the VFIB method, we make a distinction between the IB resolution, characterized by the ratio $\delta_f/D$, and the grid resolution, characterized by the ratio $D/\Delta x$, although the two are connected. We consider three levels of resolution of the immersed boundary:
\begin{enumerate*}[(i)]
  \item coarse resolution with $\delta_f/D=1/6$,
  \item medium resolution with $\delta_f/D=1/12$, and
  \item high resolution with $\delta_f/D=1/24$.
\end{enumerate*}
Unless noted otherwise, the grid spacing is chosen such that  $\delta_f/\Delta x=4$. This choice has the benefits of providing sufficient resolution of the filter kernel while retaining compact stencils for interpolations and extrapolations that use only 5 grid points in each direction. With fixed $\delta_f/\Delta x=4$, the coarse, medium, and fine IB resolutions yield increasingly fine grids characterized by the ratios $D/\Delta x=24$, 48, and 96.

\begin{figure}\centering
	\begin{subfigure}{0.9\linewidth}
	\includegraphics[width=\linewidth,trim={0 0 33ex 0},clip]{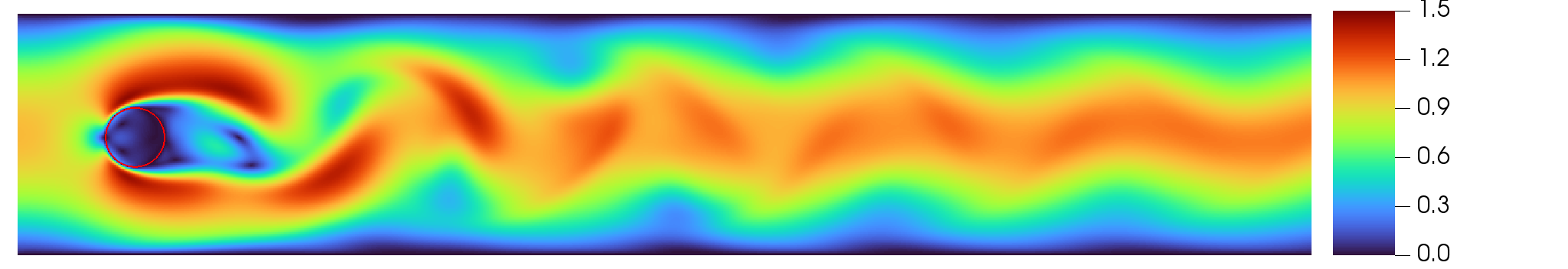}
	\caption{}
	\label{fig:cylinder_channel_visit_a}
	\end{subfigure}
	\begin{subfigure}{0.9\linewidth}
	\includegraphics[width=\linewidth,trim={0 0 33ex 0},clip]{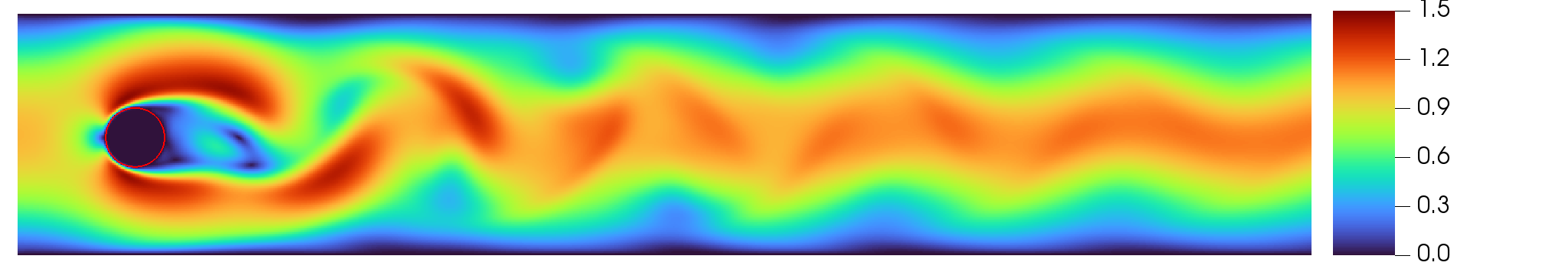}
	\caption{}
	\label{fig:cylinder_channel_visit_b}
	\end{subfigure}
	\caption{Isocontours of (a) $||\bm{u}||/U$ and (b) $||\alpha_1\overline{\bm{u}}_1||/U$ at the resolution $\delta_f/D=1/12$ and using the triangle filter kernel. The location of the immersed boundary is identified by the isolevel $\alpha_1=\alpha_2=0.5$. 	Multiplying by the volume fraction of the external fluid $\alpha_1$ hides the internal flow.}
	\label{fig:cylinder_channel_visit}
\end{figure}

Figure \ref{fig:cylinder_channel_visit_a} shows isocontours of normalized velocity magnitude $||\bm{u}||/U$ at the medium IB resolution $\delta_f/D=1/12$ obtained using the triangle filter kernel. The wake behind the cylinder shows a pair of attached eddies that form behind the cylinder similar to the observations made by \citet{trittonExperimentsFlowCircular1959} and \citet{zdravkovichSmokeObservationsFormation1969}. Figure \ref{fig:cylinder_channel_visit_a} also shows the existence of an internal flow within the cylinder. This internal flow is anticipated in the one-phase formulation because $\bm{u}=\alpha_1\overline{\bm{u}}_1+\alpha_2\overline{\bm{u}}_2$ represents the sum of both  internal and external flows. The former flow can be hidden by multiplying the total fluid velocity with the external fluid volume fraction $\alpha_1$, as shown in figure \ref{fig:cylinder_channel_visit_b}, since $\alpha_1\bm{u}\simeq \alpha_1\overline{\bm{u}}_1$.

\begin{figure}\centering
	\begin{subfigure}{0.495\linewidth}
	\includegraphics[width=\linewidth]{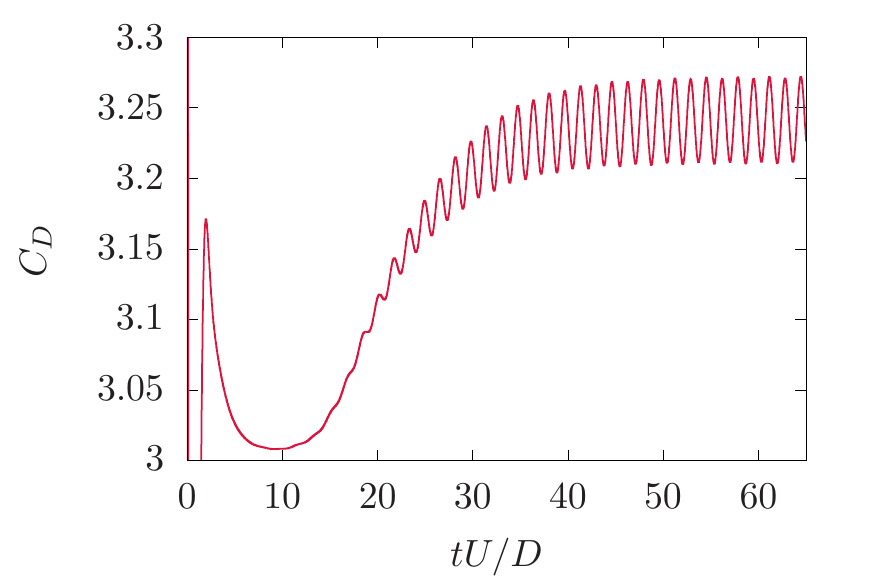}
	\caption{}
	\label{fig:triangle_channel_cd}
	\end{subfigure}
	\begin{subfigure}{0.495\linewidth}
	\includegraphics[width=\linewidth]{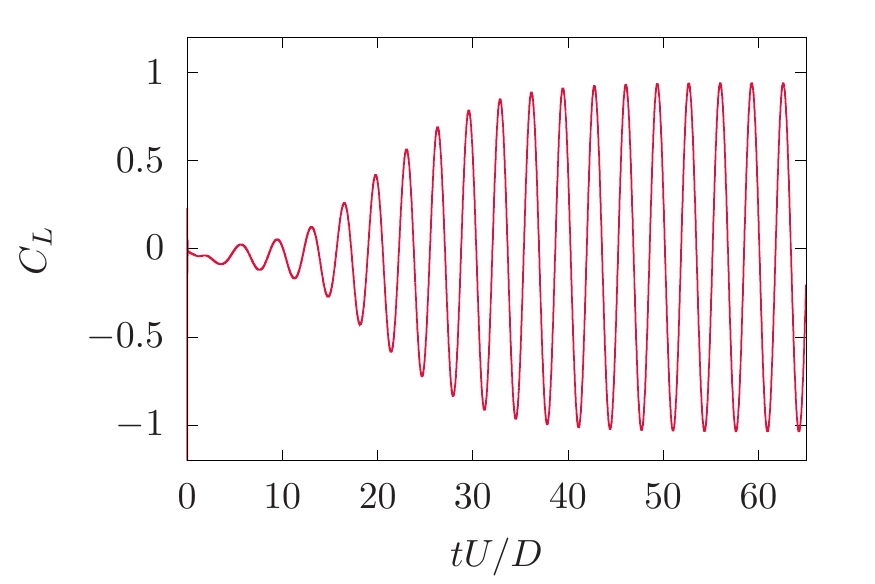}
	\caption{}
	\label{fig:triangle_channel_cl}
	\end{subfigure}
	\caption{Time evolution of the (a) drag coefficient and (b) lift coefficient for the case  of a cylinder asymmetrically placed in a channel at $\Rey_D=100$. The data is obtained using the triangle filter kernel with width $\delta_f/D= 1/24$.}
	\label{fig:channel_graphs}
\end{figure}

To assess quantitatively the performance of the VFIB method, we compare with the benchmark study of  \citet{schaferBenchmarkComputationsLaminar1996} where the results from several simulations using body-fitted meshes are complied. For the present case at $\Rey_D=100$, \citet{schaferBenchmarkComputationsLaminar1996} give a Strouhal number $\mathrm{St}=0.3\pm 0.005$, maximum drag coefficient $C_{D,\mathrm{max}}=3.23\pm 0.01$, and maximum lift coefficient $C_{L,\mathrm{max}}=1.0\pm0.01$. In the present work, drag and lift forces on the immersed solid are computed using the contribution of the external fluid only as explained in \S \ref{sec:appendix_force}.
Figure \ref{fig:channel_graphs} shows the evolution of the drag and lift coefficients at the resolution $\delta_f/D=1/24$ using the triangle kernel. These quantities reach a stationary state after $t\sim 40D/U$. We compute statics using data from this point until $t\sim 65D/U$ (about 15 periods). For the case shown in figure \ref{fig:channel_graphs}, we find $\mathrm{St}=0.300$, $C_{D,\mathrm{max}}=3.27$, and $C_{L,\mathrm{max}}=0.94$. These values are in excellent agreement with the results of body-fitted simulations reported by \citet{schaferBenchmarkComputationsLaminar1996}.

\begin{figure}\centering
  \begin{subfigure}{0.45\linewidth}
    \includegraphics[width=\linewidth]{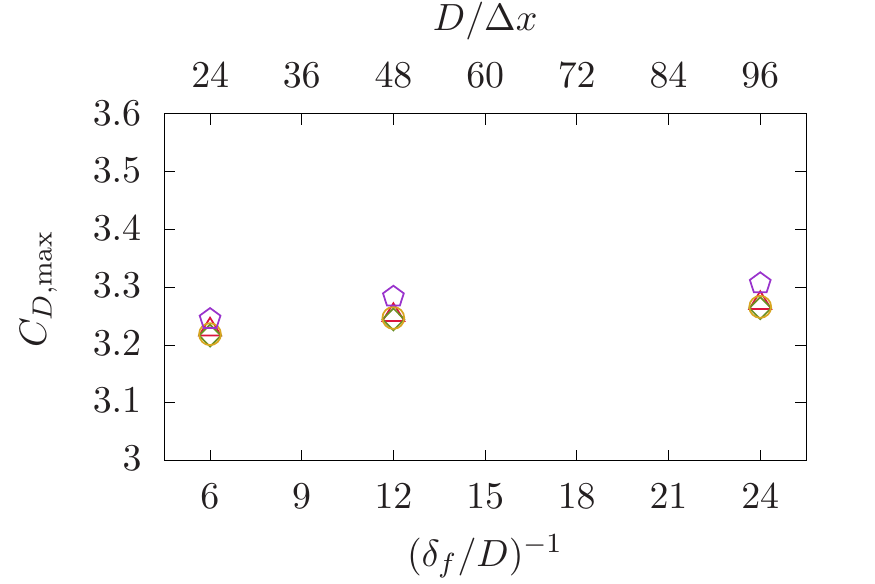}
    \caption{\label{fig:A_type_a}}
  \end{subfigure}
  \begin{subfigure}{0.45\linewidth}
    \includegraphics[width=\linewidth]{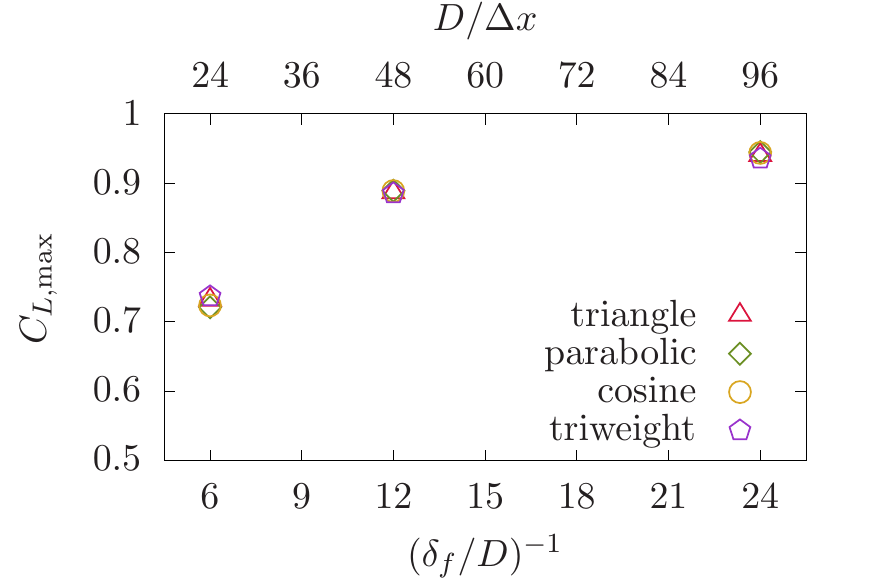}
    \caption{\label{fig:A_type_b}}
  \end{subfigure}
  \begin{subfigure}{0.45\linewidth}
    \includegraphics[width=\linewidth]{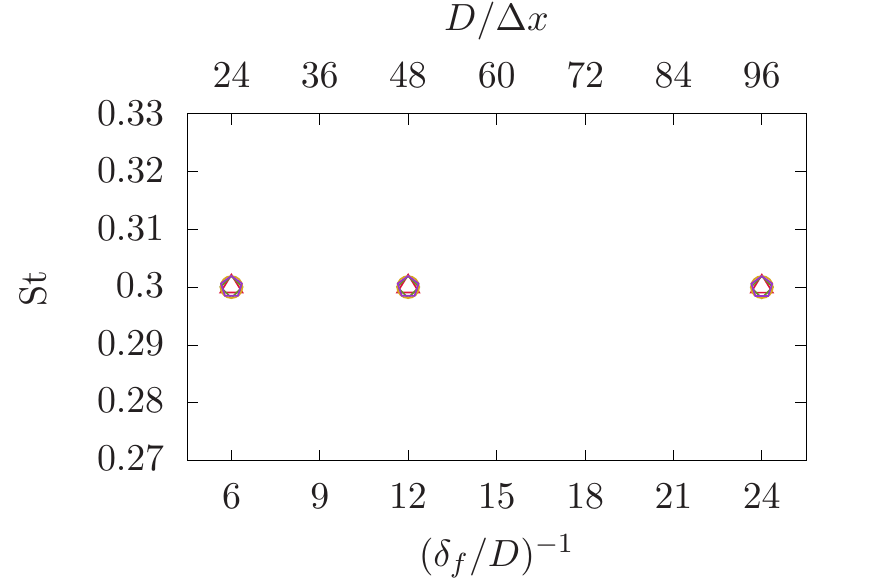}
    \caption{\label{fig:A_type_c}}
  \end{subfigure}
	\caption{Effect of varying resolution and filter kernels on the (a) maximum drag coefficient, (b) maximum lift coefficient, and (c) Strouhal number for the case of a cylinder asymmetrically placed in a channel at $\Rey_D=100$. \label{fig:A_filter}}
\end{figure}

Figure \ref{fig:A_filter} shows that increasing the IB resolution causes $C_{D,\mathrm{max}}$, $C_{L,\mathrm{max}}$, and the Strouhal number $\mathrm{St}$ to converge to the benchmark values, regardless of the choice of filter kernel.  Note that with improving IB resolution, i.e., decreasing ratio $\delta_f/D$, more grid points are required to maintain the ratio $\delta_f/\Delta x=4$. From figure \ref{fig:A_filter}, we observe that the predicted drag, lift, and Strouhal converge to the benchmark values as $\delta_f/D$ is reduced for all filter kernels considered. The choice of filter kernel has little effect on the predicted maximum drag and lift coefficients, and Strouhal number provided that the immersed boundary is sufficiently well resolved.

\begin{figure}\centering
  \begin{subfigure}{0.45\linewidth}
    \includegraphics[width=\linewidth]{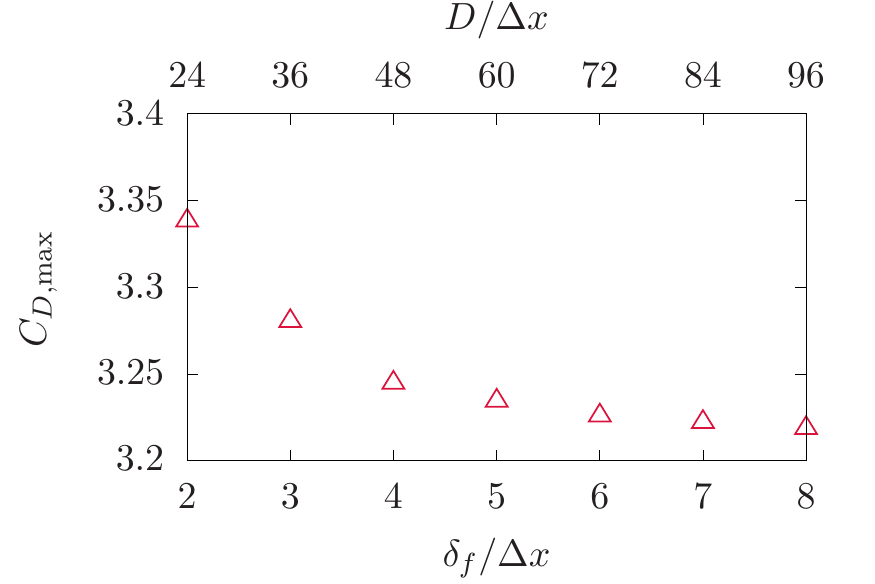}
    \caption{\label{fig:A_ratio_a}}
  \end{subfigure}
  \begin{subfigure}{0.45\linewidth}
    \includegraphics[width=\linewidth]{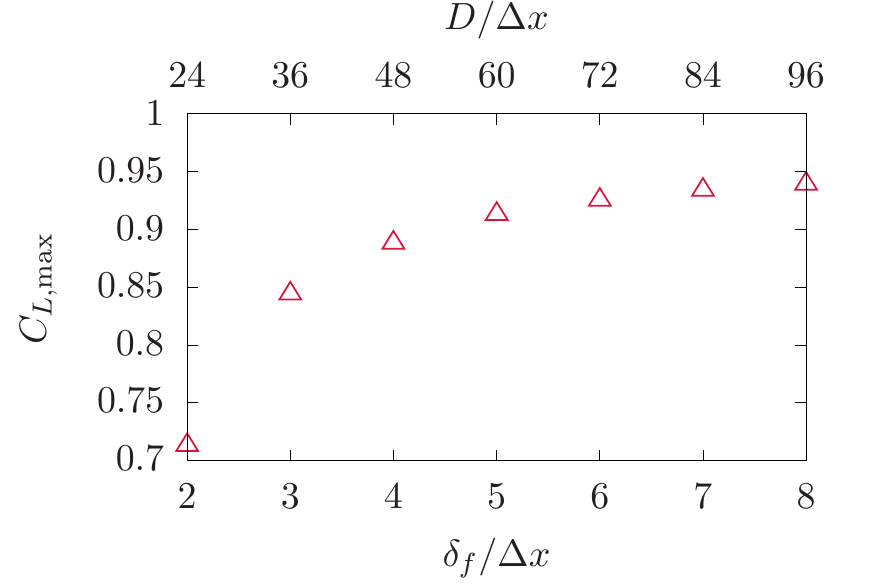}
    \caption{\label{fig:A_ratio_b}}
  \end{subfigure}
   \caption{Effect of increasing the grid resolution at fixed filter width $\delta_f=D/12$ on the (a) maximum drag coefficient, (b) maximum lift coefficient on a cylinder asymetrically placed in a channel at $\Rey_D=100$. For these simulations, the triangle filter kernel is used. Results are well converged for $\delta_f/\Delta x\geq 4$.\label{fig:A_ratio}}
\end{figure}

For a fixed filter width $\delta_f$, the predicted drag and lift coefficients also converge with decreasing $\Delta x$.
In figure \ref{fig:A_ratio}, the filter width is maintained at $\delta_f/D=1/12$, while the grid spacing $\Delta x$ progressively reduced such that the ratio $\delta_f/\Delta x$ varies from 2 to 8. At the lower end $\delta_f/\Delta x=2$, the interpolation/extrapolation require stencils with only 3 grid points in each direction, thus, making this choice the most computationally efficient one. At $\delta_f/\Delta x=8$, the stencils require 9 grid points in each direction. As shown in figure \ref{fig:A_ratio}, $C_{D,\mathrm{max}}$, $C_{L,\mathrm{max}}$ converge with increasing ratio $\delta_f/\Delta x$. The choice of $\delta_f/\Delta x=4$ is a happy medium between small stencils and well converged results.

\subsection{Fixed cylinder in free stream}
\label{sssec:freestream}

We now consider the case of a fixed cylinder of diameter $D$ placed in free stream.
{\color{revision} Unless otherwise noted, the simulations are carried out in a computational domain of size $L_x = L_y = 26D$, similar to the configuration considered by \citet{uhlmannImmersedBoundaryMethod2005} . A uniform inflow with  velocity $U$ is prescribed at the inlet $x=0$. The cylinder is placed at $x=6D$ and centered in the y direction. The parameters are chosen such that the Reynolds number $\Rey_D=UD/\nu$ is 100. We also consider an additional simulation in an enlarged domain of size $L_x = L_y = 40D$ to compare with the sharp-interface IB method of \citet{mittalVersatileSharpInterface2008}.}

\begin{figure}\centering
	\begin{subfigure}{\linewidth}\centering
    \includegraphics[width=0.9\linewidth]{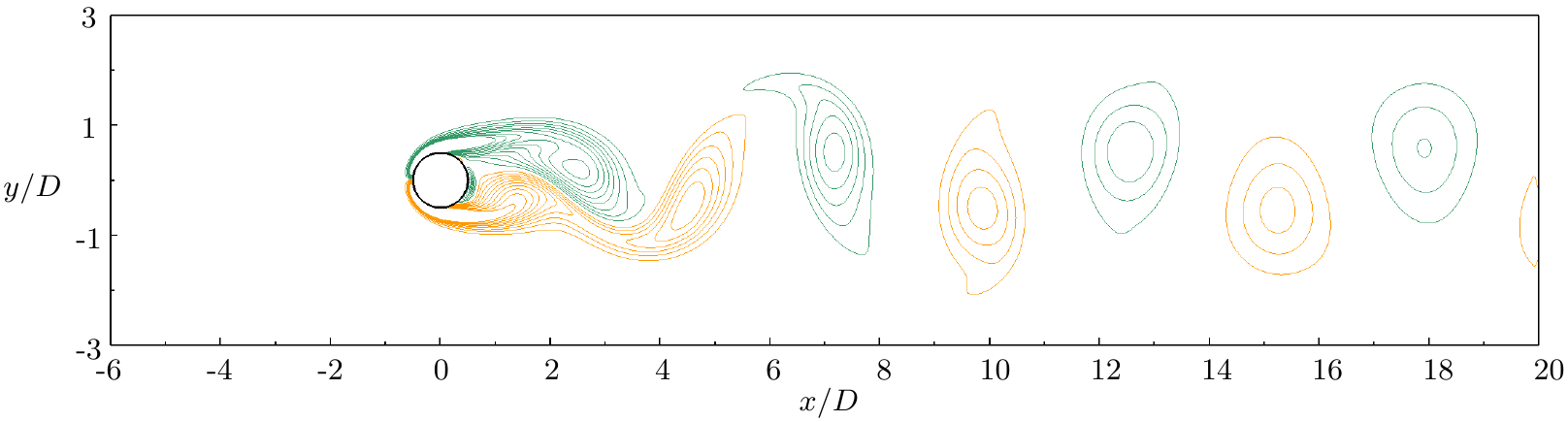}
    \caption{\label{fig:freestream_shedding}}
	\end{subfigure}
	\begin{subfigure}{0.45\linewidth}
	\includegraphics[width=\linewidth]{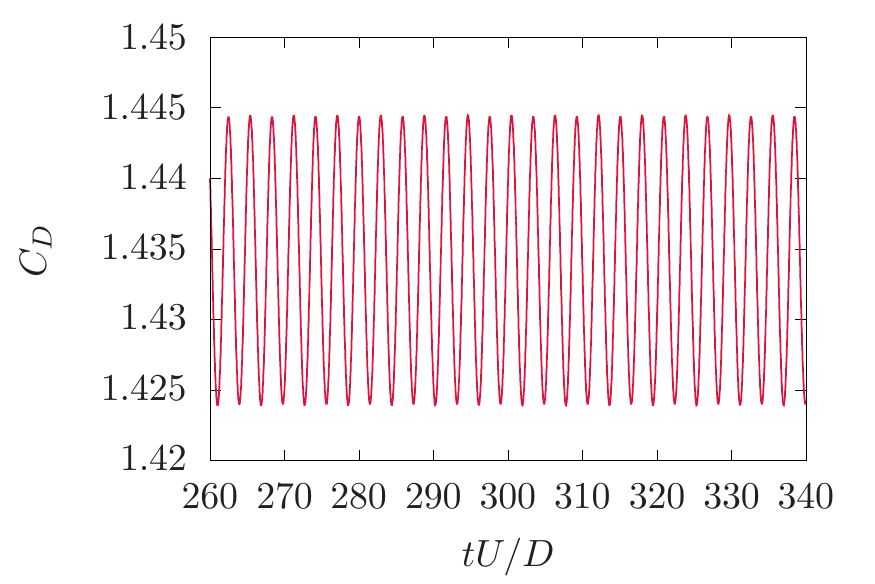}
	\caption{\label{fig:freestream_cd}}
	\end{subfigure}
	\begin{subfigure}{0.45\linewidth}
	\includegraphics[width=\linewidth]{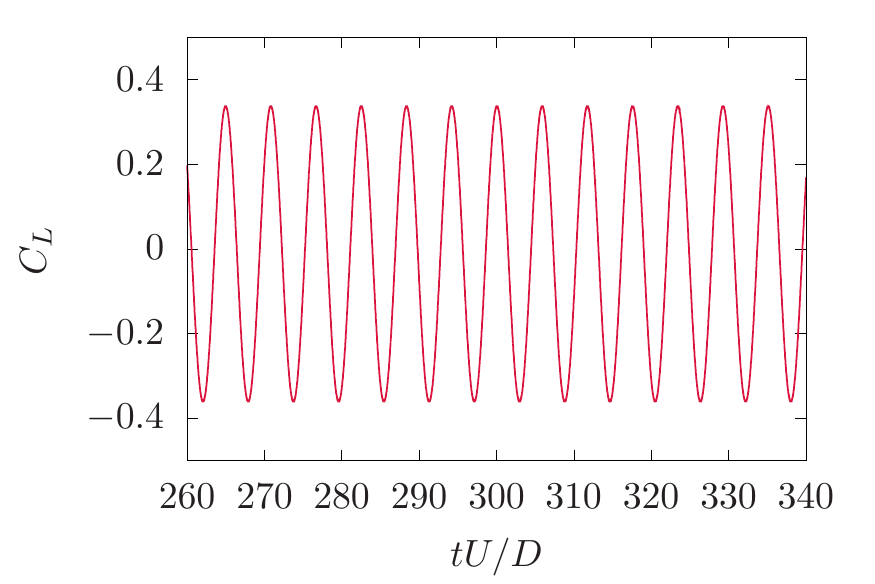}
	\caption{\label{fig:freestream_cl}}
	\end{subfigure}
	\caption{Flow past an immersed cylinder at $\Rey_D=100$ obtained using the VFIB method with the triangle filter kernel and width $\delta_f=D/24$. Vortex shedding can be seen from (a) alternating isolevels of positive (orange) and negative (green) vorticity. This results in fluctuating (b) drag and (c) lift coefficients.\label{fig:freestream}}
\end{figure}

Once the flow over the immersed cylinder is established, a vortex street is observed in the wake of the cylinder. Figure \ref{fig:freestream_shedding} shows that vortices are shed from the top and bottom of the cylinder at alternating intervals. This leads to oscillating drag and lift forces as shown in figures \ref{fig:freestream_cd} and \ref{fig:freestream_cl}. The periodic vortex shedding is characterized by the Strouhal number $\mathrm{St}= f_0D/U$ where $f_0$ is the shedding frequency extracted from the lift force.

\begin{table}\centering\color{revision}
  \def\refa{\small\citet{uhlmannImmersedBoundaryMethod2005}}
  \def\refb{\small\citet{liuPreconditionedMultigridMethods1998}}
	\def\refc{\small\citet{mittalVersatileSharpInterface2008}}
  \caption{\color{revision} Characteristics of drag and lift coefficients for a cylinder in a freestream at $\mathrm{Re}_D = 100$. The results from the VFIB method are compared to body-fitted simulations of \citet{liuPreconditionedMultigridMethods1998} and the immersed boundary approach of \citet{uhlmannImmersedBoundaryMethod2005} and \citet{mittalVersatileSharpInterface2008}. \label{tab:cylinder_freestream}}
  \begin{tabularx}{\linewidth}{bssssss}
  \hline
           & $\delta_f/D$ & $D/\Delta x$ & $\overline{C}_D$ & $C'_D$ & $C'_L$ & $\mathrm{St}$\\\hline
 \refb     & --           & --           & 1.350            & $0.012$& $0.339$& $0.165$ \\\hline
 Present   & $1/3$        & $10$         & $1.364$          & $0.003$& $0.214$& $0.165$\\
           & $1/5$        & $20$         & $1.424$          & $0.008$& $0.320$& $0.165$\\
           & $1/10$       & $40$         & $1.434$          & $0.010$& $0.348$& $0.165$\\
 Present (enlarged) & $1/16$ & $64$			 & $1.355$  		 		& $0.010$& $0.331$& $0.165$\\\hline
 \refa     & --           & $38.4$       & $1.501$          & $0.011$& $0.349$& $0.172$\\\hline
 \refc     & --           & $66$         & $1.350$          & --     & --     & $0.165$\\\hline
\end{tabularx}
\end{table}
\textcolor{revision}{ Table \ref{tab:cylinder_freestream} contains a summary of the Strouhal number $\mathrm{St}$, mean drag coefficient $\overline{C}_D$, fluctuating drag coefficient $C'_D$, and fluctuating lift coefficient is $C'_L$ from the present simulations with the triangle filter kernel. Data obtained with the body-fitted mesh simulations of \citet{liuPreconditionedMultigridMethods1998}, and IB methods of \citet{uhlmannImmersedBoundaryMethod2005} and \citet{mittalVersatileSharpInterface2008} are also included for comparison. For the case shown in figure \ref{fig:freestream}, with resolution $\delta_f=D/24$, we find 
 $\mathrm{St}=0.165$, $\overline{C}_D=1.434$, $C'_D=0.010$, and $C'_L=0.346$. For comparison, \citet{liuPreconditionedMultigridMethods1998} give $\mathrm{St}=0.165$, $\overline{C}_D=1.35$,  $C'_D=0.012$, and $C'_L=0.339$ from simulations with body-fitted mesh, whereas \citet{uhlmannImmersedBoundaryMethod2005} gives $\mathrm{St}=0.172$, $\overline{C}_D=1.501$,  $C'_D=0.011$, and $C'_L=0.349$ using his immersed boundary method. Thus, the agreement with the benchmark results of \citet{liuPreconditionedMultigridMethods1998} is improved using the VFIB method, particularly for the Strouhal number and mean drag coefficient. The latter is within 6.2\% of the value reported by \citet{liuPreconditionedMultigridMethods1998}, compared to an over-prediction by 11.1\% given by \citet{uhlmannImmersedBoundaryMethod2005}.}

\textcolor{revision}{The drag over-prediction reduces further to less than 0.37\% when the simulation is carried in the enlarged domain and at the higher resolution $\delta_f/D = 1/16$. This resolution, which corresponds to $D/\Delta x = 64$, and enlarged domain are chosen to approximately match the domain and resolution used by \citet{mittalVersatileSharpInterface2008}, that is $D/\Delta x = 66$. As shown in table \ref{tab:cylinder_freestream}, our results match very closely those of \citet{mittalVersatileSharpInterface2008} and the reference results of \citet{liuPreconditionedMultigridMethods1998}.}

\begin{figure}\centering
	\begin{subfigure}{0.45\linewidth}
	\includegraphics[width=\linewidth]{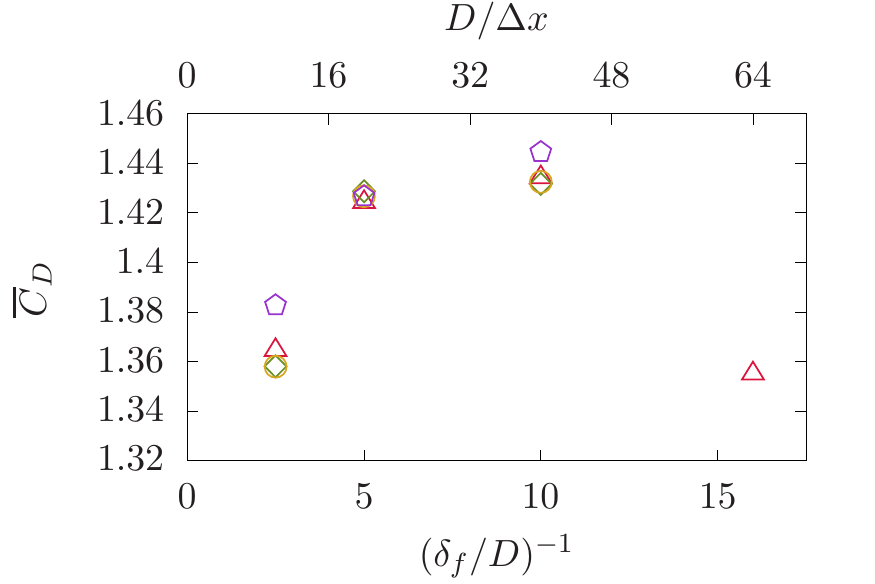}
	\caption{}
	\end{subfigure}
	\begin{subfigure}{0.45\linewidth}
	\includegraphics[width=\linewidth]{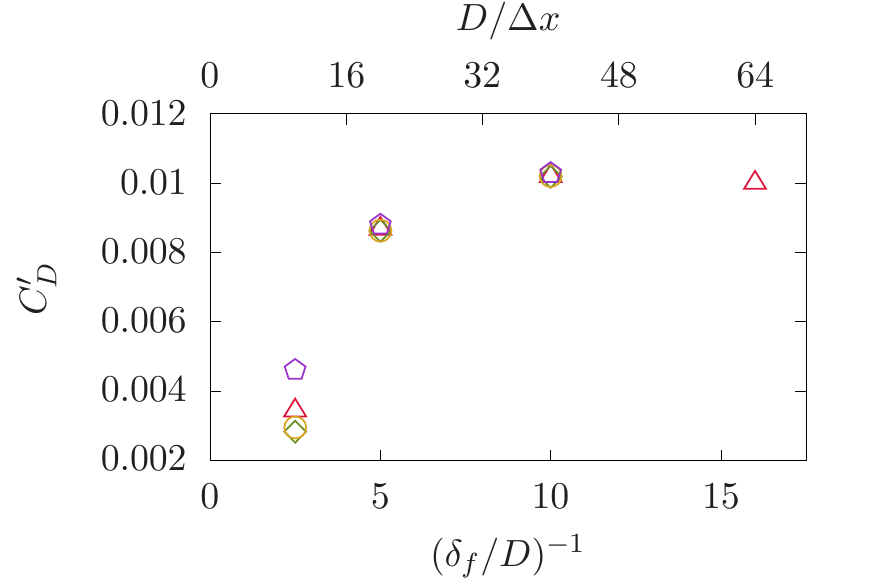}
	\caption{}
	\end{subfigure}\\
	\begin{subfigure}{0.45\linewidth}
	\includegraphics[width=\linewidth]{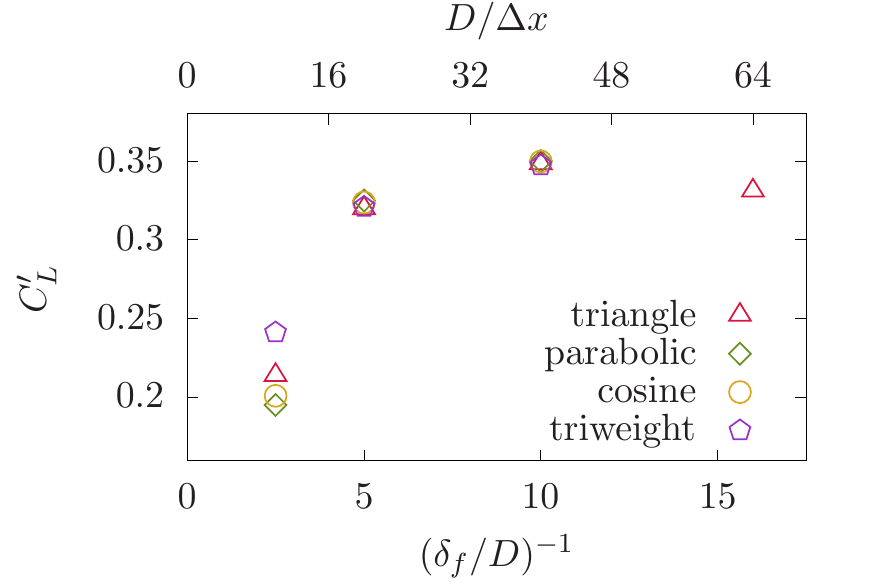}
	\caption{}
	\end{subfigure}
	\begin{subfigure}{0.45\linewidth}
	\includegraphics[width=\linewidth]{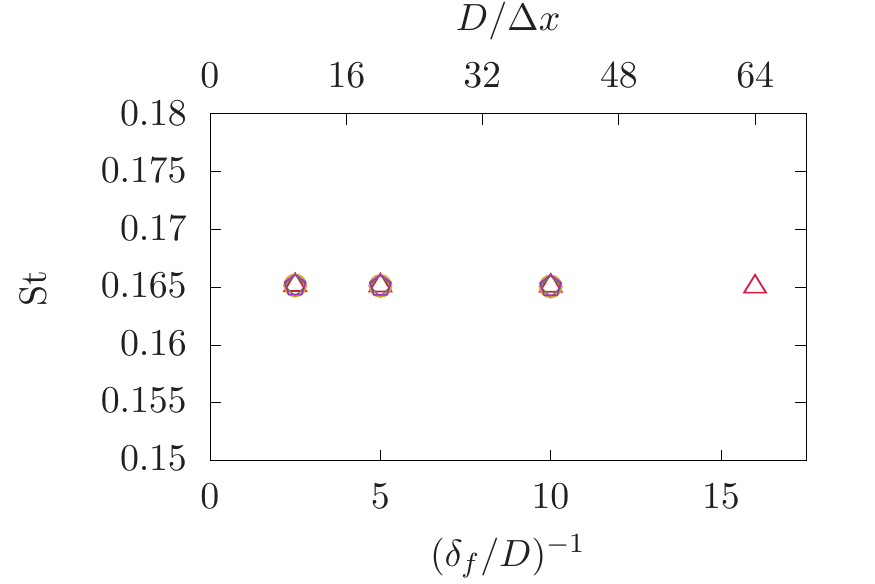}
	\caption{}
	\end{subfigure}
	\caption{Effect of varying resolution and filter kernel on the (a) mean drag, (b) drag fluctuation, (c) lift fluctuation, and (d) Strouhal number for the case of an immersed cylinder at $\Rey_D=100$. \label{fig:B_filter}}
\end{figure}
Figure \ref{fig:B_filter} shows the effect of varying the filter kernel and width. Similar to what we have shown in the previous numerical example, these results show that the predicted $\mathrm{St}$, $\overline{C}_D$,  $C'_D$, and $C'_L$ converge with improving resolution of the immersed cylinder and that the choice of the filter kernel has little impact if the immersed boundary is well resolved.

\subsection{\textcolor{revision}{Laterally} oscillating cylinder in a uniform crossflow}
\label{sssec:oscillating_l}

In this test, we consider a \textcolor{revision}{laterally} oscillating immersed cylinder in a uniform crossflow at $\Rey_D=185$. The configuration is identical to the one described in the previous test (\S\ref{sssec:freestream}), with the difference that, now, the cylinder has forced oscillations around its position. The displacement of the cylinder center is $\Delta y_c = 0.2D\sin(2\pi f_e t)$, where $f_e=0.8f_0$ is the forced oscillation frequency, and $f_0$ is the natural shedding frequency at $\Rey_D=185$. We consider two spatial resolutions at $\delta_f/D=1/6$ (coarse), and $1/12$ (medium), and vary the time step $\Delta t$ to yield a maximum Courant-Friedrich-Levy number $\mathrm{CFL}_{\mathrm{max}}$ between 0.5 and 0.1. Since we have determined that the choice of filter kernel plays little role for well-resolved immersed boundaries, we perform tests with the triangle kernel only.

\begin{figure}\centering
	\begin{subfigure}{0.54\linewidth}
    \includegraphics[width=\linewidth]{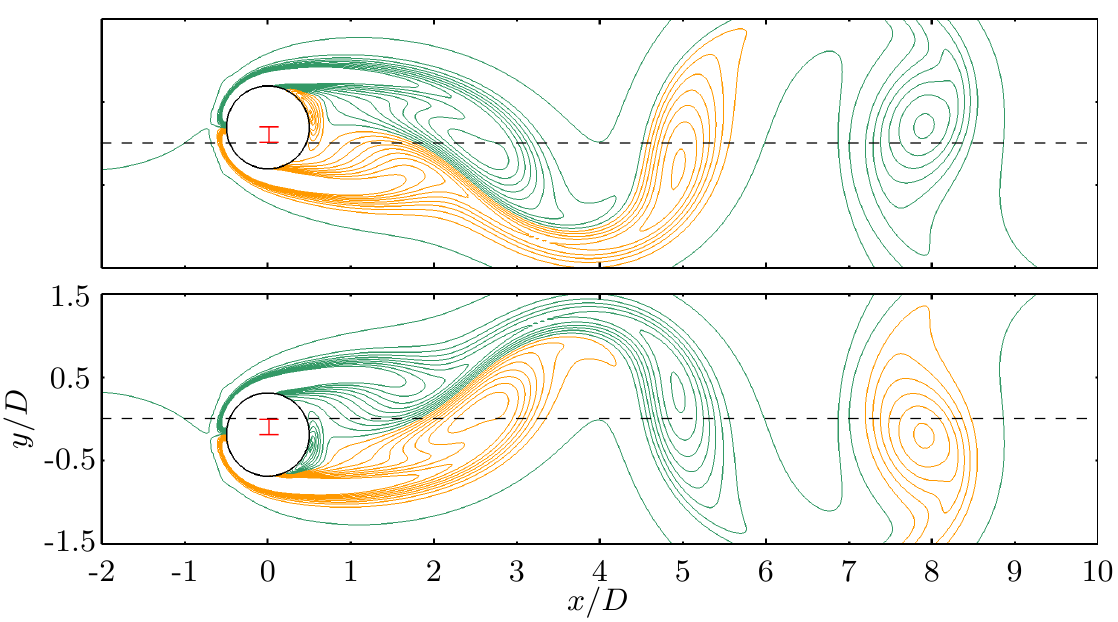}
    \caption{\label{fig:oscillating_vort}}
	\end{subfigure}
	\begin{subfigure}{0.45\linewidth}
	\includegraphics[width=\linewidth]{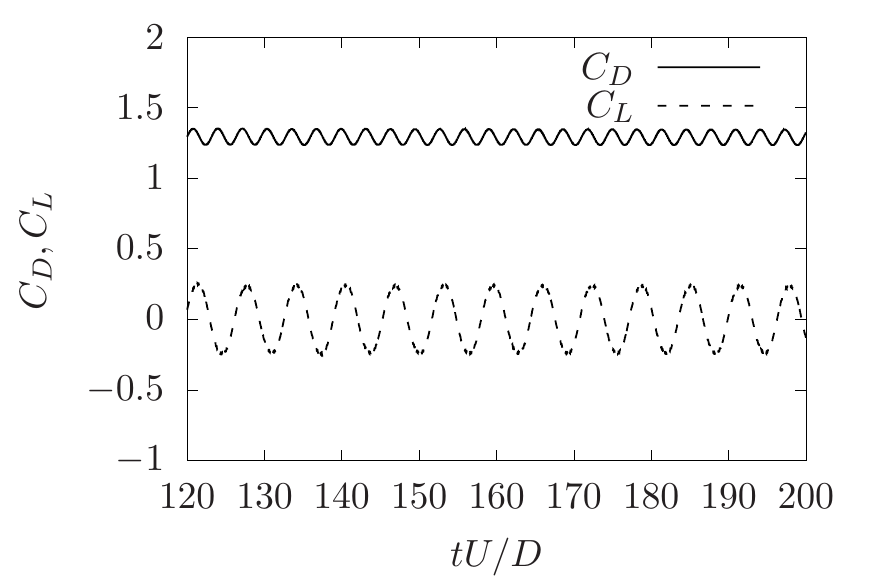}
	\caption{\label{fig:oscillating_cdcl}}
	\end{subfigure}
	\caption{Flow past an oscillating immersed cylinder with a crossflow at $\Rey_D=185$ obtained using the VFIB method with the triangle filter kernel, width $\delta_f/D=1/24$, and $\mathrm{CFL}_{\mathrm{max}} = 0.5$. (a) Isocontours of positive (orange) and negative (green) vorticity when the cylinder is at its highest (top) and lowest (bottom) positions. This results in fluctuations of (b) drag and lift coefficients.}
  \label{fig:oscillating_stat}
\end{figure}
Once the flow reaches a stationary state, a vortex street is observed in the wake of the cylinder. Figure \ref{fig:oscillating_vort} shows the vortex contours when the cylinder is at its highest and lowest displacements. Compared to the case of a static cylinder, the vortex contours from the upper end of the cylinder are elongated, and the contours form the base of the cylinder are tightened, when the cylinder is in the highest dispalcement.  This tightening of the base vorticity is due to the upward motion of the oscillating cylinder generating vorticity at the base. The opposite behavior is observed in the configuration of lowest displacement. Similar observations were also reported by \citet{luCALCULATIONTIMINGVORTEX1996} who used simulations with body-conforming meshes. As a result of this vortex shedding, oscillatory fluctuations in the coefficient of drag and lift are witnessed as shown in figure \ref{fig:oscillating_cdcl}.

Table \ref{tab:cylinder_oscillate} contains a summary of the mean drag coefficient $\overline{C}_D$, drag fluctuations $C_{D,\mathrm{rms}}$, and lift fluctuations $C_{L,\mathrm{rms}}$. The values reported by \citet{luCALCULATIONTIMINGVORTEX1996} and \citet{uhlmannImmersedBoundaryMethod2005} are also reported for comparison.

At the coarse resolution $\delta_f/D = 1/6$, there is a strong sensitivity to temporal errors, but, these reduce significantly with decreasing $\mathrm{CFL}_{\mathrm{max}}$. Taking  the values reported by \citet{luCALCULATIONTIMINGVORTEX1996} as reference, the VFIB method at $\mathrm{CFL}_{\mathrm{max}}=0.5$ yields $\overline{C}_D$, $C_{D,\mathrm{rms}}$ and $C_{L,\mathrm{rms}}$ that are within $2.5\%$, $22.8\%$, and $25\%$ respectively, of the reference values. Reducing $\mathrm{CFL}_{\mathrm{max}}$ to 0.25 or 0.1 causes deviations in predicted $\overline{C}_D$ and $C_{D,\mathrm{rms}}$ to reduce to $1\%$ and $3.5\%$, respectively. In addition, the deviations in $C_{L,\mathrm{rms}}$ reduce to $14\%$ and $1.8\%$ at $\mathrm{CFL}_{\mathrm{max}}=0.25$ and 0.1. Note that even at $\mathrm{CFL}_{\mathrm{max}}=0.5$, and the coarse resolution $\delta_f/D = 1/6$ ($D/\Delta x = 24)$, the values predicted using the VFIB method agree better with the reference values in \citep{luCALCULATIONTIMINGVORTEX1996} than those given by \citet{uhlmannImmersedBoundaryMethod2005} with $D/\Delta x=38.4$ and $\mathrm{CFL}_{\mathrm{max}}=0.6$ (see table \ref{tab:cylinder_oscillate}).

With improving resolution of the immersed boundary to $\delta_f/D=1/12$, the predicted $C_{D,\mathrm{rms}}$ and $C_{L,\mathrm{rms}}$ converge to those given by \citet{luCALCULATIONTIMINGVORTEX1996} even at $\mathrm{CFL}_{\mathrm{max}} = 0.5$. The mean drag coefficient $\overline{C}_D$ is off by $3.2\%$ at $\mathrm{CFL}_{\mathrm{max}} = 0.5$ and converges to the exact reference value at $\mathrm{CFL}_{\mathrm{max}} = 0.25$.

{\color{revision}
This greatly enhanced performance compared to the method of \citet{uhlmannImmersedBoundaryMethod2005} is the result of three improvements:
\begin{enumerate*}[(i)]
  \item a more accurate way of computing the hydrodynamic force using equation (\ref{eq:53}),
  \item the fact that, in Uhlmann's method, the surface markers have a Lagrangian volume $\Delta V_m=\Delta x^3$ which, as we discussed at the end of section 3.1, is not accurate, and
  \item significantly reduced spurious force oscillations, as we discuss next.
\end{enumerate*}
}

\begin{table}\centering
  \def\refa{\small\citet{uhlmannImmersedBoundaryMethod2005}}
  \def\refb{\small\citet{luCALCULATIONTIMINGVORTEX1996}}
  \def\refc{\small\citet{yangSmoothingTechniqueDiscrete2009}}
\caption{Characteristics of drag and lift on a transversely oscillating cylinder at $\Rey_D = 185$. Results from the present VFIB method are compared to results of \citet{luCALCULATIONTIMINGVORTEX1996} obtained using body-confirming grid, and the results with the immersed boundary method of \citet{uhlmannImmersedBoundaryMethod2005} and smoothing technique of \citet{yangSmoothingTechniqueDiscrete2009}. For the latter, we report only their smoothest case obtained using their smoothed 4-point piecewise Dirac delta. \label{tab:cylinder_oscillate}}
  \begin{tabularx}{\linewidth}{bssssss}
  \hline
           & $\delta_f/D$ & $D/\Delta x$ & $\mathrm{CFL}_{\mathrm{max}}$ & $\overline{C}_D$ & $C_{D,\mathrm{rms}}$ & $C_{L,\mathrm{rms}}$\\\hline
 \refb     & --           & --           & --                            & $1.25$           & $0.040$              & $0.18$ \\\hline
 Present   & $1/6$        & $24$         & $0.5$                         & $1.28$           & $0.031$              & $0.135$\\
           & $1/6$        & $24$         & $0.25$                        & $1.24$           & $0.039$              & $0.155$\\
           & $1/6$        & $24$         & $0.1$                         & $1.24$           & $0.039$              & $0.177$\\
           & $1/12$       & $48$         & $0.5$                         & $1.29$           & $0.040$              & $0.18$ \\
           & $1/12$       & $48$         & $0.25$                        & $1.25$           & $0.040$              & $0.18$ \\\hline
 \refa     & --           & $38.4$       & $0.6$                         & $1.380$          & $0.045$              & $0.176$\\\hline
 \refc     & --           & $50$         & --                            & $1.29$           & $0.043$              & $0.07$ \\\hline
\end{tabularx}
\end{table}

\begin{figure}\centering
	\begin{subfigure}{0.45\linewidth}
    \includegraphics[width=\linewidth]{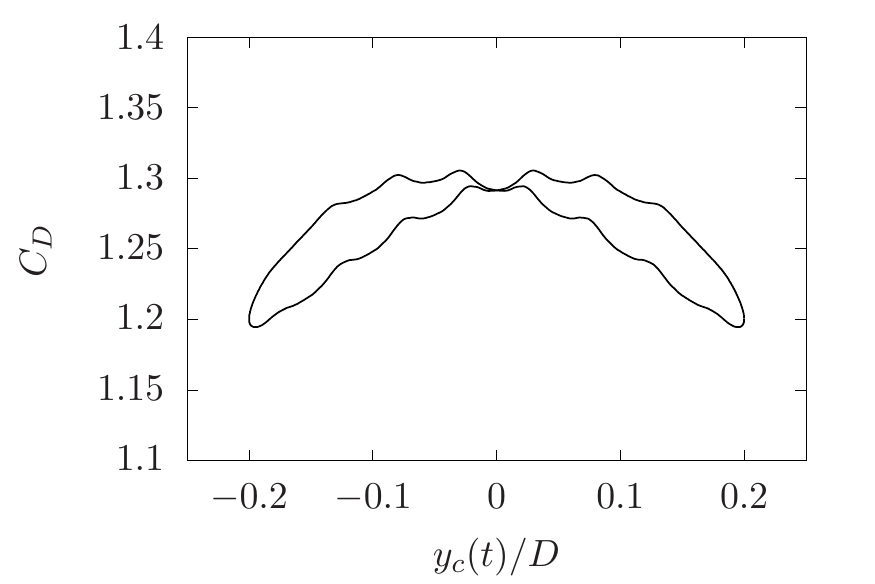}
    \caption{\label{fig:cdyc_ddx24}}
	\end{subfigure}
	\begin{subfigure}{0.45\linewidth}
	\includegraphics[width=\linewidth]{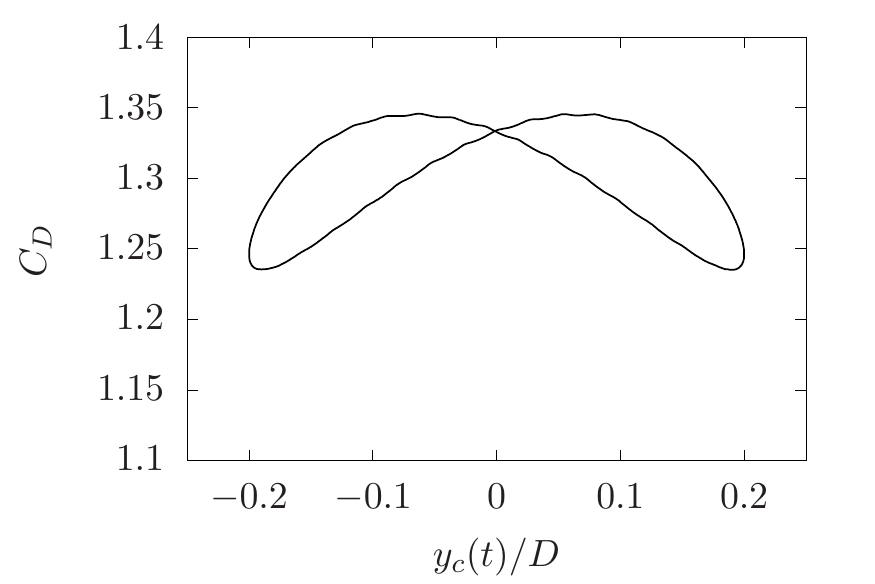}
	\caption{\label{fig:cdyc_ddx48}}
	\end{subfigure}
	\caption{Time-periodic variation of $C_D$ with respect to the position of the oscillating cylinder for $\Rey_D=185$ at (a) $\mathrm{CFL}_{\mathrm{max}} = 0.5$, $\delta_f/D=1/6$ ($D/\Delta x = 24$) and (b) $\mathrm{CFL}_{\mathrm{max}} = 0.5$, $\delta_f/D=1/12$ ($D/\Delta x = 48$).}
  \label{fig:cdyc_oscillate}
\end{figure}

Here, we emphasize that our treatment of extrapolations using analytical calculations of the integrals shown in \S \ref{sec:appendix_interp} reduces \textcolor{revision}{spurious force fluctuations considerably} without the need for any smoothing technique like the one developed by \citet{yangSmoothingTechniqueDiscrete2009}.  Figure \ref{fig:cdyc_oscillate} shows a graph of $C_D$ during 1 period of oscillation at $\mathrm{CFL}_{\mathrm{max}} = 0.5$ at the resolutions $\delta_f/D=1/6$ and $1/12$.
Similar to what has been reported by \citet{yangSmoothingTechniqueDiscrete2009} and \citet{uhlmannImmersedBoundaryMethod2005}, spurious force fluctuations can be seen in the graph with the coarse resolution $\delta_f/D=1/6$ shown in figure \ref{fig:cdyc_ddx24}. These oscillations have length scale comparable to the mesh spacing $\Delta x$, and depend largely on how interpolations and extrapolations are performed numerically. Increasing the spatial resolution reduces the amplitude of the spurious oscillations significantly. Without using any smoothing technique, the curve in figure \ref{fig:cdyc_ddx48} ($\delta_f/D=1/12$) is as smooth as the one shown in \citep{yangSmoothingTechniqueDiscrete2009} obtained with similar numerical parameters. This is because, as the immersed solid moves, cells that are only partially covered by the filter kernel are still forced when the extrapolation is performed as shown in \S \ref{sec:appendix_interp}, even if this coverage is minute. In contrast, in methods based on numerical approximations of the integrals using the mid-point rule, a cell would not be forced until at least half of it is covered by the kernel (or, similarly, the support of the discretized Dirac delta), which leads to jagged and discontinuous forcing as the immersed solid moves.

\subsection{Flow past a sphere}
\label{sssec:sphere}
We now consider a three dimensional case involving the flow past a sphere. This flow has been studied in detail by numerous investigators, thus, providing a wealth of information to validate the present immersed boundary strategy. Notably, \citet{johnsonFlowSphereReynolds1999a} used a numerical approach based on a body-fitted grid to resolve the steady and unsteady dynamics associated with the transition from attached wake to vortex shedding in the wake of a sphere up to Reynolds number of 300. \citet{johnsonFlowSphereReynolds1999a} showed that the wake is steady and axisymmetric up to $\Rey_D\simeq200$. For $\Rey_D$ between 210 and 270, the wake remains steady although the axisymmetry is lost. Values of $\Rey_D$ greater than 270 lead to periodic vortex shedding in the form of a sequence of hairpin vortices.

\begin{figure}\centering
    \includegraphics[width=4.8in]{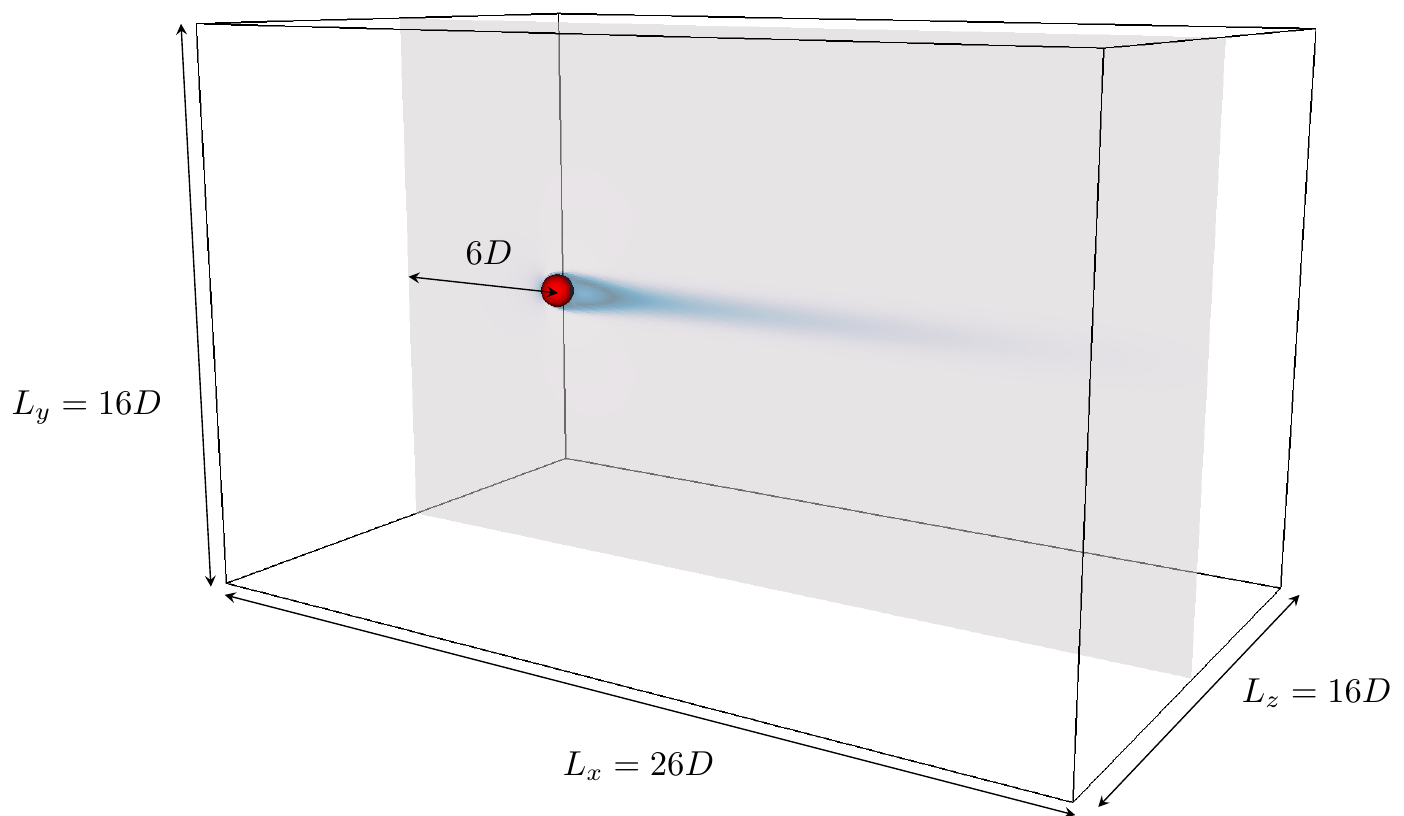}
    \caption{Computational domain used for simulations of flow past a sphere.\label{fig:sphere_domain}}
\end{figure}
In order to compare with the results of \citet{johnsonFlowSphereReynolds1999a}, we perform simulations of the flow over a sphere at Reynolds numbers from $\Rey_D=25$ to 300. The simulation domain extends by $L_x = 26D$ in the flow direction, and $L_z=L_y=16D$ in the two normal directions. Figure \ref{fig:sphere_domain} shows a schematic of the configuration. Since we have established that the choice of filter kernel has little effect effect in well-resolved simulations, we consider simulations with the triangle filter kernel only.

\begin{figure}\centering
    \includegraphics[width=4.8in]{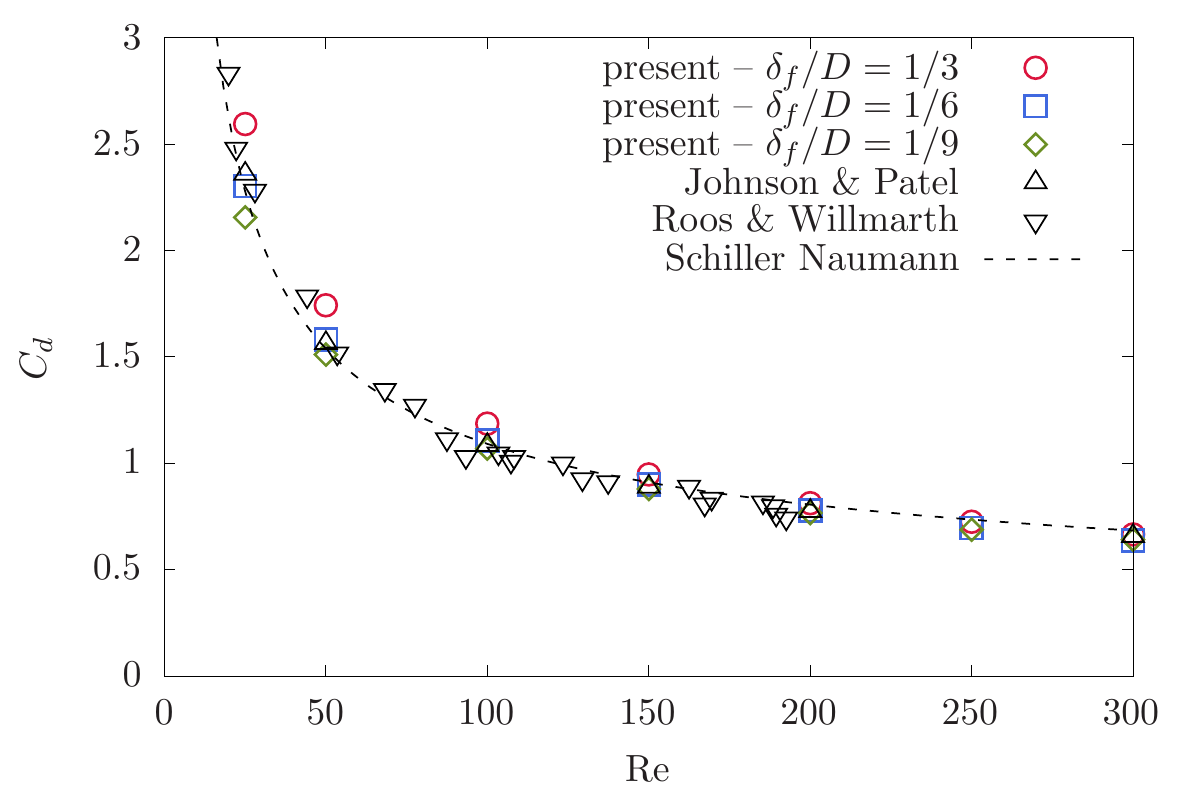}
    \caption{Variation of the drag coefficient with Reynolds number for the flow past a sphere.\label{fig:sphere_CD}}
\end{figure}
Figure \ref{fig:sphere_CD} shows the variation of the drag coefficient with Reynolds number. For comparison, data obtained using the VFIB method is plotted alongside results of the body-fitted grid simulations of \citet{johnsonFlowSphereReynolds1999a}, and experiments of \citep{roosExperimentalResultsSphere1971}. Further, the Schiller-Naumann correlation $C_D^{SN}=\frac{24}{Re_p}(1+0.15 Re_{p}^{0.687})$ is also reported on the same plot \citep{schillerDragCoefficientCorrelation1933}. At low Reynolds numbers, the low resolution simulations with $\delta_f/D=1/3$ give a drag coefficient in excess by up to 10\% of the Schiller-Naumann drag coefficient, but still within good agreement with other numerical and experimental data. Increasing the resolution of the immersed boundary to $\delta_f/D=1/6$ and $\delta_f/D=1/9$ causes the predicted drag values to converge. At the highest resolution, there is excellent agreement with prior data and the Schiller-Naumann correlation.

\begin{figure}\centering
    \begin{subfigure}{1\linewidth}\centering
     \includegraphics[width=\linewidth]{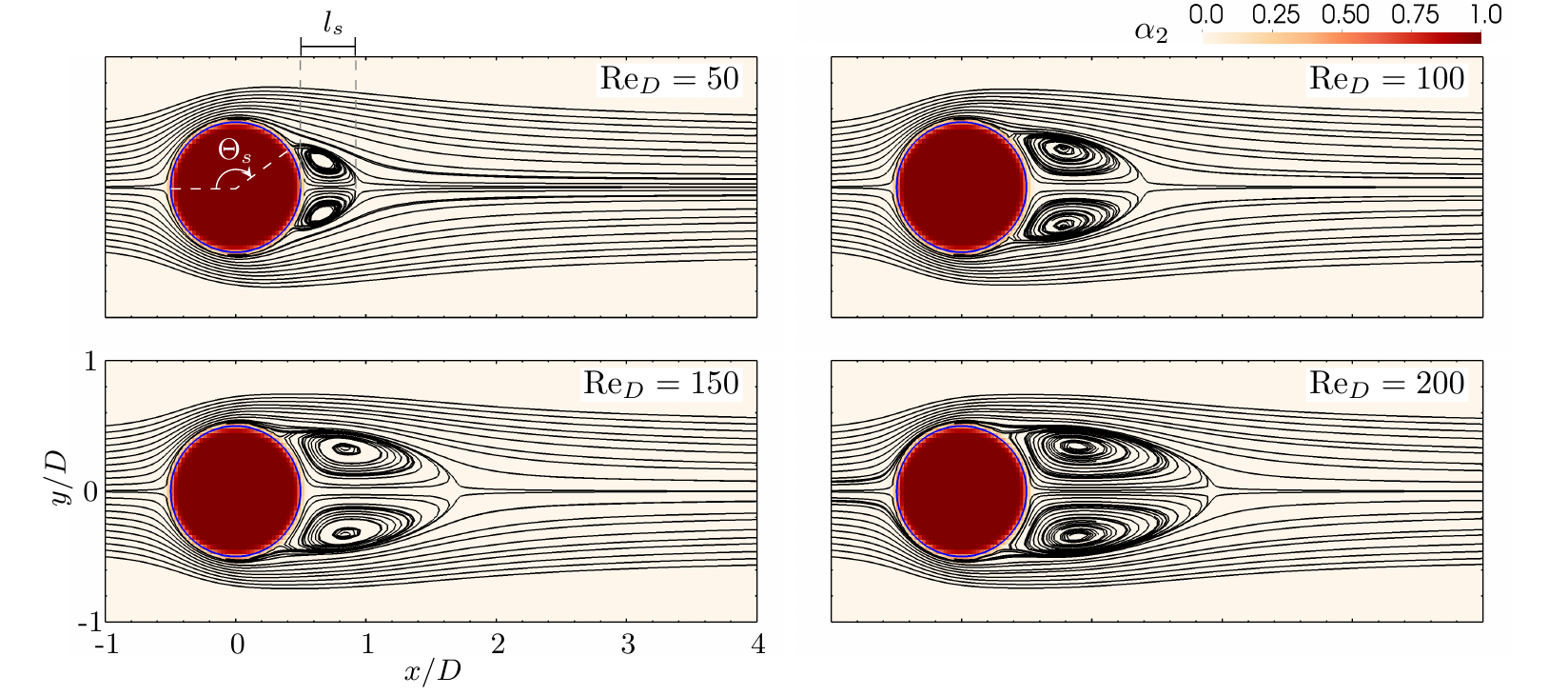}
     \caption{\label{fig:sphere_streamlines}}
    \end{subfigure}
    \begin{subfigure}{0.49\linewidth}\centering
      \includegraphics[width=3.3in]{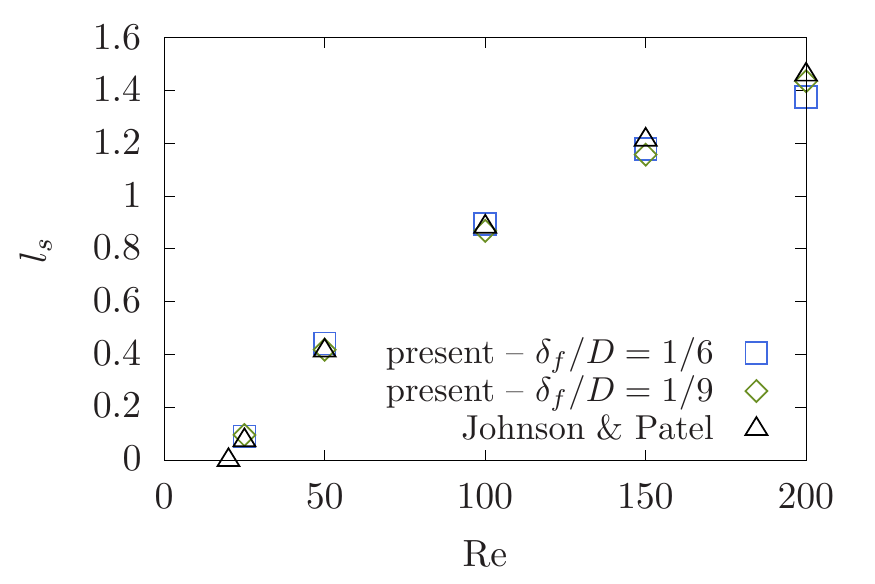}
	  \caption{\label{fig:separation_length}}
    \end{subfigure}
    \begin{subfigure}{0.49\linewidth}\centering
      \includegraphics[width=3.3in]{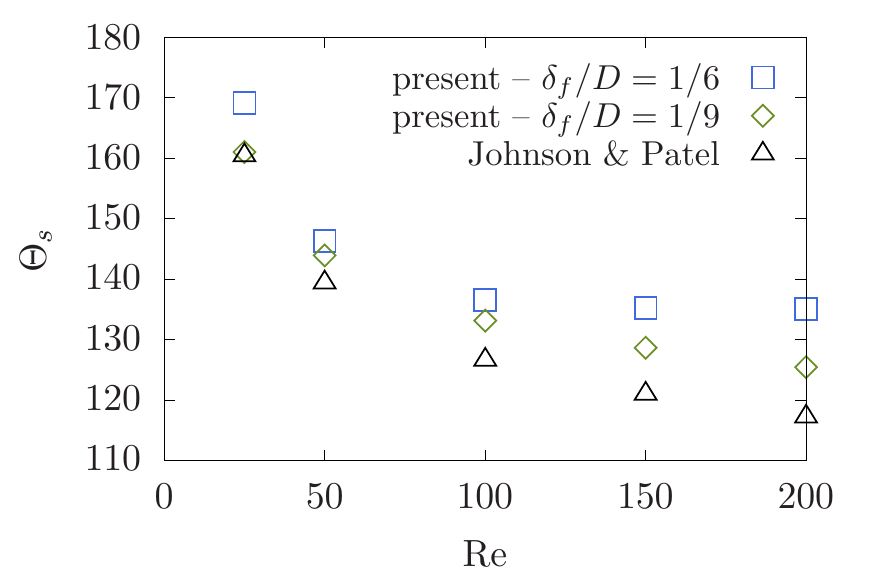}
  	  \caption{\label{fig:separation_angle}}
	\end{subfigure}
   \caption{Characteristics of the wake behind an immersed sphere in the axisymmetric regime: (a) streamlines at $\Rey_D=50$, $100$, $150$, and $200$, (b) length of the separation bubble, and (b) separation angle. \label{fig:sphere_Re_up_to_200}}
\end{figure}
In figure \ref{fig:sphere_Re_up_to_200}, we examine the characteristics of the wake up to $\Rey_D=200$. Figure \ref{fig:sphere_streamlines} shows the streamlines in the $x$-$y$ plane going through the sphere's center for the four cases at  $\Rey_D=50$, 100, 150, and 200, and at the resolution $\delta_f/D=1/9$. At these Reynolds numbers, the flow is steady and axisymmetric. The near-wake flow exhibits a recirculation bubble similar to previously reported observations \citep{tanedaExperimentalInvestigationWake1956,pruppacherRelationsDragFlow1970,johnsonFlowSphereReynolds1999a}. As shown in figure \ref{fig:sphere_streamlines}, flow separation happens at an angle $\Theta_s$ that decreases with increasing Reynolds number, while the bubble length $l_s$ increases with increasing Reynolds number. In figures \ref{fig:separation_length} and \ref{fig:separation_angle}, we compare the values of bubble length $l_s$ and separation angle $\Theta_s$ obtained with the VFIB method with those obtained by \citet{johnsonFlowSphereReynolds1999a}. Figure \ref{fig:separation_length} shows that the resolution $\delta_f/D=1/6$ is sufficient to yield excellent agreement with the data of \citet{johnsonFlowSphereReynolds1999a}. However, capturing the separation angle $\Theta_s$ accurately requires higher resolution. For the cases with $\delta_f/D=1/6$, we find a separation angle that is larger than the previously reported values by about 8 degrees at $\Rey_D=25$, and 18 degrees at $\Rey_D=200$. These deviations reduce to 1 degree at $\Rey_D=25$, and 8 degrees at $\Rey_D=200$, when the resolution is increased to $\delta_f/D=1/9$. The trend in figure \ref{fig:separation_angle} shows that the angles obtained using the VFIB method converge towards the reference values, but higher resolution is required to achieve agreement within a few degrees.

\begin{figure}\centering
   \includegraphics[width=\linewidth]{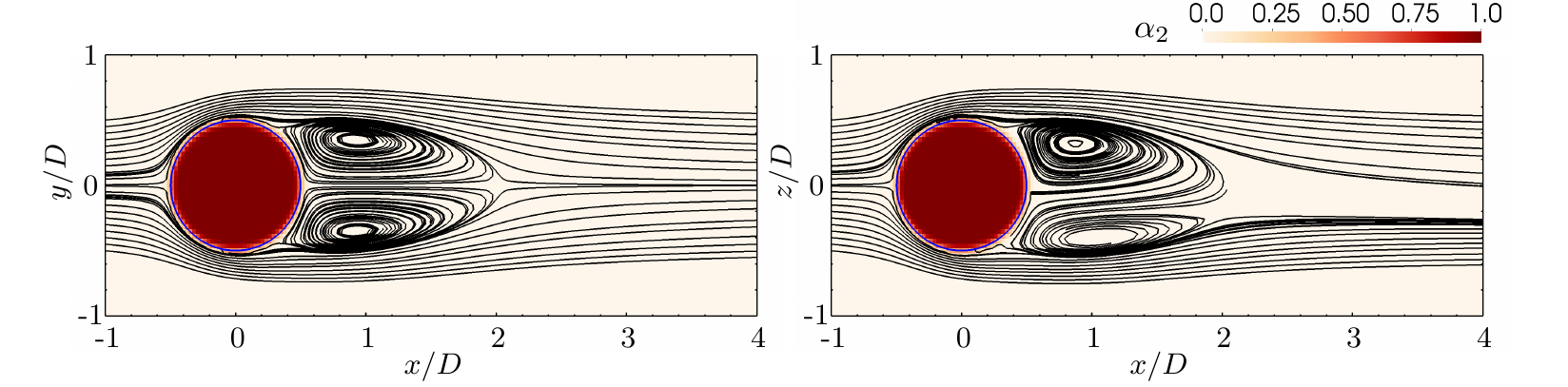}
   \caption{Streamlines showing the asymmetric near-wake flow past a sphere at $\Rey_D=250$. \label{fig:streamlines_Re250}}
\end{figure}
When the Reynolds number is increased to $\Rey_D=250$, the VFIB method correctly captures the transition to an asymmetric near-wake. Figure \ref{fig:streamlines_Re250} shows flow streamlines at $\Rey_D=250$ and resolution $\delta_f/D=1/9$ in the $x$-$y$ and $x$-$z$ planes. While the flow remains steady, an azimuthal mode appears in the near wake. This mode is symmetric about the $x$-$z$ plane. These observations are in agreement with the results of \citet{johnsonFlowSphereReynolds1999a} who also observe loss of axisymmetry due to the emergence of an azimuthal mode with planar symmetry similar to what is shown in figure \ref{fig:streamlines_Re250}.

%\begin{figure}\centering
%   \includegraphics[width=0.7\linewidth]{figures/fig17.jpeg}
%   \caption{Vortex shedding in the wake of a sphere at $\Rey_D=300$. \label{fig:wake}}
%\end{figure}

\subsection{\textcolor{revision}{Freely falling sphere under gravity}}
\label{sssec:freely_falling}

{\color{revision}
In this final test, we consider the case of a freely falling sphere under gravity. The particle position $\bm{x}_p$, velocity $\bm{u}_p$, and angular velocity $\bm{\omega}_p$ are updated at each time step by solving the following equations of motion,
\begin{eqnarray}
  \frac{d\bm{x}_p}{dt}&=&\bm{u}_p \label{eq:particle_1}\\
  m_p\frac{d\bm{u}_p}{dt} &=& \iint_{S_I}\bm{n}\cdot\bm{\tau}_1dS + (\rho_p-\rho_f)\frac{\pi D^3}{6}\bm{g}_v \label{eq:particle_2}\\
  I_p\frac{d\bm{\omega}_p}{dt} &=& \iint_{S_I}(\bm{y}-\bm{y}_p)\times\bm{n}\cdot\bm{\tau}_1dS
  \label{eq:particle_3}
\end{eqnarray}
where $\rho_p$, $D$,  $m_p=\rho_p(\pi/6)D^3$, and $I_p=m_p D^2/10$ are the particle density, diameter, mass, and moment of inertia. Here, $\bm{g}_v$ denotes the gravitational acceleration. The first term on the right hand side of (\ref{eq:particle_2}) represents the hydrodynamic stresses exerted by the external fluid and computed using the expression (\ref{eq:53}). This requires an update of the volume fraction field $\alpha_f$ at each step, which is carried out as described in section \ref{sec:vol_frac}.

We use the experiments from \citet{mordantVelocityMeasurementSettling2000} as benchmark. In these experiments, a spherical particle is released with zero velocity in a fluid initially at rest. The particle accelerates until it reaches its terminal velocity. The latter is controlled by two non-dimensional numbers: the density ratio $\rho_p/\rho_f$ and the Galileo number $\mathrm{Ga}=\sqrt{(\rho_p/\rho_f-1)g_v D^3}/\nu$. 

For our comparison, we match case 2 from \citet{mordantVelocityMeasurementSettling2000} for which $\rho_p/\rho_f=2.56$ and $\mathrm{Ga}=255.35$. The simulations are carried out in a domain with dimensions $L_x = L_z = 8D$, and $L_y = 40D$ in the direction of gravity. Three filter sizes are considered: $\delta_f=D/2$, $\delta_f=D/3$, and $\delta_f=D/4$. With the requirement $\delta_f/\Delta x=4$, this yields 8, 12, and 16 grid points across the particle diameter, respectively. In all these runs, $\mathrm{CFL}_\mathrm{max}\sim 0.35$.

\begin{figure}\centering
   	\includegraphics[width=5in]{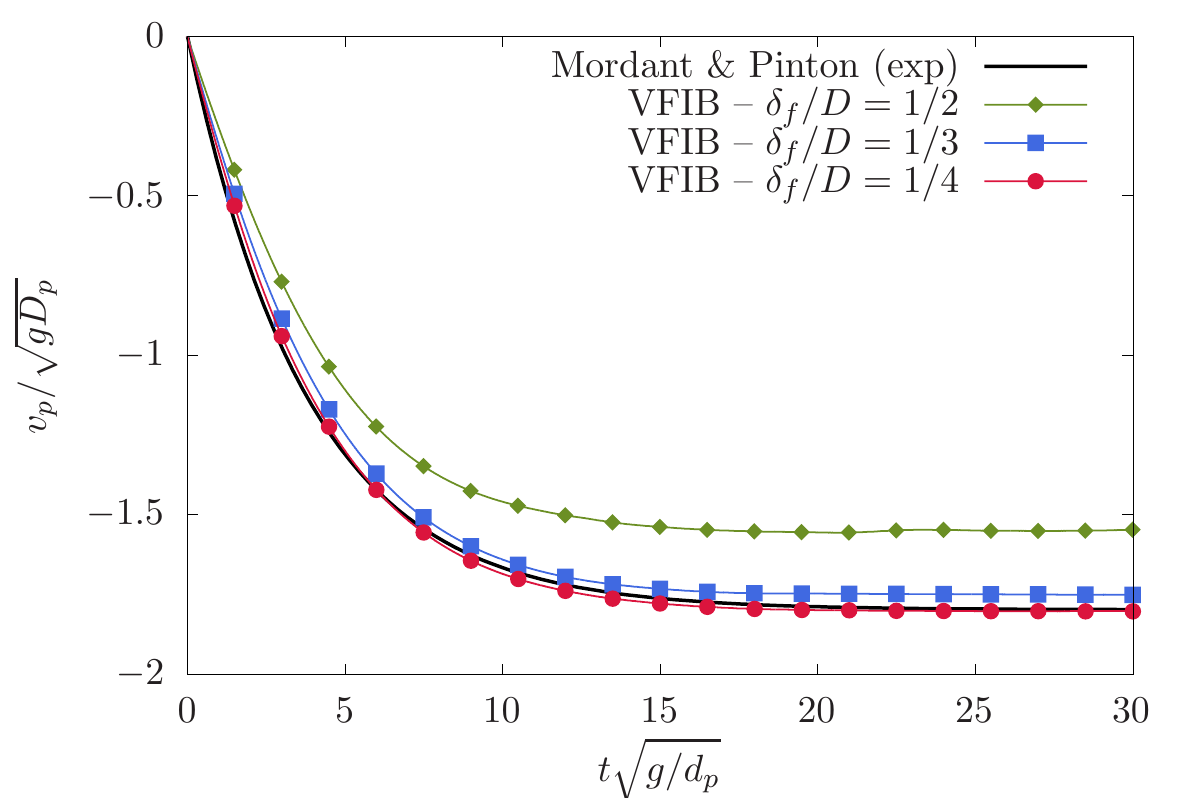}	  
	\caption{Time series of the particle settling velocity for the case with $\rho_p/\rho_f=2.56$ and $\mathrm{Ga}=41798$. The solid line is obtained using the experimental correlation of \citet{mordantVelocityMeasurementSettling2000}. The Reynolds number and Froude number based on the terminal velocity are $\Rey=367$ and $\mathrm{Fr}=1.80$, respectively.}
	\label{fig:settling}
\end{figure}
Figure \ref{fig:settling} shows comparison of the particle velocity time series with the experimentally obtained data by \citet{mordantVelocityMeasurementSettling2000}. As can be seen from the figure, the case with resolution $\delta_f/D=1/4$ shows excellent agreement with the experiment.  The difference between the computed terminal velocity and the one reported by \citet{mordantVelocityMeasurementSettling2000} is less than 0.3\%. The Reynolds number and Froude number based on the terminal velocity are $\Rey=v_{p,t}d_p/\nu=368.39$ and $\mathrm{Fr}=v_{p,t}/\sqrt{gD}=1.80$, respectively, in the simulations and $\Rey=367.41$ and $\mathrm{Fr}=1.80$ in the experiments. There is good agreement even for the case with resolution  $\delta_f/D=1/3$ where the relative error for the settling velocity is about 2.5\%. In the coarsest case $\delta_f/D=1/2$, the relative error increases to about 13.8\%.
}

\section{Conclusion}
\label{sec:conclusion}

In this paper, we have presented a novel immersed boundary method, called VFIB, derived by applying the volume-filtering technique of  \citet{andersonFluidMechanicalDescription1967a}. Without assuming any discretization, we obtain filtered transport equations where the effect of the immersed boundary appears as a forcing on the right-hand side of these equations. This new framework can be regarded as a generalization of previous immersed boundary methods, and provides a theoretical footing for further extensions. We provided extensive details on how this method can be implemented in existing flow solvers and showed that it yields excellent results in two- and three-dimensional cases with static, \textcolor{revision}{forcibly moving, and freely moving} immersed boundaries. 

\textcolor{revision}{We shall emphasize that there are no restrictions on the type of computations that can be performed with the VFIB method. As shown in the numerical examples, any moving or static can be accurately represented and accounted for including for problems in fluid-structure interaction, impeller/turbine setups, and Particle-Resolved DNS of particle-laden flows.}

The present work contains key innovations that enable us to answer several open questions in the literature of immersed boundary methods. \textcolor{revision}{First, by volume-filtering the original Navier-Stokes equations, we derived \emph{analytically} the immersed boundary forcing term. This is in contrast to the majority of prior methods that build the IB forcing based on numerical considerations, such as requiring that cells coinciding with the IB have velocities that match the IB velocity at the next step \citep{uhlmannImmersedBoundaryMethod2005,leeSourcesSpuriousForce2011,seoSharpinterfaceImmersedBoundary2011,schneidersAccurateMovingBoundary2013,luoFullscaleSolutionsParticleladen2007,breugemSecondorderAccurateImmersed2012,kempeImprovedImmersedBoundary2012,tschisgaleNoniterativeImmersedBoundary2017}. A major aspect of our method is the fact that we intentionally take into account the finite width of the filter. This allows us to carry out the analytical derivation, find an expression for the IB forcing, derive equations for the fluid/solid volume fraction, and forces exerted on the IB by the external flow.}
% able to obtain the coupling term at the interface that arise out of the surface integrals as part of the volume filtering process. While almost all previous approaches couple the fluid-solid phases using an artificial force term added to the momentum transport equation , we provide a formulation that gives physical meaning to the forcing term between the phases.}

Second, using the VFIB method, we are able to elucidate the role of the internal flow obtained when the IB forcing is limited to the solid-fluid boundary only.  As we show in the one-phase formulation, this flow has physical meaning and exists because the immersed object is in reality hollow, has infinitely thin shell, and is filled with identical fluid to the one on the outside. It follows that when the immersed object moves, the fluid inside is also affected, resulting in additional stresses on the boundary. In order to isolate the hydrodynamic force due to the external fluid only, the contribution due to the internal fluid must be removed. The details are given in \S \ref{sec:appendix_force}. This approach removes the need for ad-hoc fixes, like artificially retracting the immersed boundary, to get accurate hydrodynamic forces.

Third, we showed that the volumes associated with Lagrangian markers depend on the local topology of the interface and can be determined simply using the smearing length $\ell$ and a triangular tessellation of the interface ($\Delta V_m=\ell(\bm{x}_m)A_m$). These Lagrangian volumes arise after discretization of the surface integrals accounting for the IB forcing. At a point $\bm{x}_m$ on the interface, the smearing length is the inverse of the surface density at that point $\ell(\bm{x}_m)=\Sigma(\bm{x}_m)^{-1}$. Its value depends on the local curvature, and choice of filter kernel, and is generally on the order of the filter width $\delta_f$.

Fourth, we provided an efficient procedure to compute the solid volume fractions $\alpha_s$. The approach is based on solving a Poisson equation for $\alpha_s$ rather than using the cumbersome and computationally expensive definition. Computing the volume fraction allows efficient tagging of interior/exterior cells, and is used in the procedure to isolate the hydrodynamic stresses due to the external fluid from the total stresses due to both internal and external fluids.

Fifth, we showed a path forward to extend the VFIB method to Large Eddy Simulations with immersed boundaries. By rigorously filtering the Navier-Stokes equations, we showed the presence of sub-filter scale terms. While we have not considered this in the present study, closures for these terms can be carried out using existing LES models, or derived by applying coarse filters to highly-resolved simulations with the VFIB method.

Lastly, we found that the choice of filter kernel does not impact the solution considerably, provided that the immersed boundary is well resolved. Among the four kernels we have considered, the triangle kernel is the simplest to implement and is, thus, the preferred one. The resolution of the immersed boundary depends on the ratio $\delta_f/\delta_c$ where $\delta_c$ is the characteristic corrugation scale of the interface. In two-dimensional cases of flow past a cylinder of diameter $D$, we found little difference between the filter kernels when $\delta_f/D\leq 1/12$. With increasing resolution, drag and lift coefficients converge towards the reference values to within one or two percent. In three dimensional cases of flow past a sphere, we get accurate drag coefficient, separation angle, and recirculation bubble length with a resolution $\delta_f/D=1/9$.

\section*{Acknowledgement}
The authors acknowledge support from the US National Science Foundation (award \#2028617, CBET-FD). Computing resources were provided by ACCESS allocation PHY200082 and Research Computing at Arizona State University.

\appendix
{\color{revision}
\section{Alternative form of the volume-filtered equations}\label{sec:multiphase_form}

Here, we explain how equations (\ref{eq:twophase_1}) and (\ref{eq:twophase_2}) can be further transformed to get the familiar form used in multiphase flows. Starting from equation (\ref{eq:twophase_2}), we expand the first term on the right hand side to obtain
\begin{eqnarray}
  \rho_f\left(\frac{\partial}{\partial t}(\alpha_f\overline{\bm{u}}_f)+\nabla\cdot(\alpha_f\overline{\bm{u}}_f\,\overline{\bm{u}}_f)\right)&=&-\alpha_f \nabla \overline{p} + \alpha_f\nabla \cdot\left(\mu_f\left(\nabla\overline{\bm{u}}_f+\nabla\overline{\bm{u}}_f^T-\frac{2}{3}(\nabla\cdot\overline{\bm{u}}_f)\bm{I}\right)+\bm{R}_{\mu,f}\right)   	-\widetilde{\bm{F}}_{I,f}-\nabla \cdot (\alpha_f\bm{\tau}_\mathrm{sfs,f}).\nonumber\\[-1ex]
\end{eqnarray}
In this usual form, the term $\widetilde{\bm{F}}_{I,f}$ is given by
\begin{eqnarray}
  \widetilde{\bm{F}}_{I,f} &=&  \bm{F}_{I,f} - \nabla\alpha_f\cdot \overline{\bm{\tau}},\label{eq:appendix_1}
\end{eqnarray}
where $\overline{\bm{\tau}}=-\overline{p}\bm{I} +\mu_f (\nabla \overline{\bm{u}}_f+\nabla \overline{\bm{u}}_f^T-2/3(\nabla\cdot\overline{\bm{u}}_f)\bm{I}+\bm{R}_\mu$ is the filtered stress tensor. Inserting $\overline{\bm{\tau}}$ in identity (\ref{eq:identity2}), we can show
\begin{eqnarray}
  \nabla\alpha_f\cdot \overline{\overline{\bm{\tau}}}-\iint_{\bm{y}\in S_I}\bm{n} \cdot\overline{\bm{\tau}}g(\bm{x}-\bm{y})dS&=&\alpha_f\left(\overline{\nabla\cdot\overline{\bm{\tau}}}-\nabla\cdot\overline{\overline{\bm{\tau}}}\right)= O(\delta_f^2). \label{eq:appendix_2}
\end{eqnarray}
Equation (\ref{eq:appendix_2}) combined with the fact that $\overline{\overline{\bm{\tau}}}=\overline{\bm{\tau}}+O(\delta_f^2)$ leads to
\begin{eqnarray}
  \widetilde{\bm{F}}_{I,f} = \iint_{\bm{y}\in S_I}\bm{n} \cdot\left(\bm{\tau}-\overline{\bm{\tau}}\right)g(\bm{x}-\bm{y})dS+O(\delta_f^2). 
\end{eqnarray}
The momentum equation is then written in the following form
\begin{eqnarray}
  \rho_f\left(\frac{\partial}{\partial t}(\alpha_f\overline{\bm{u}}_f)+\nabla\cdot(\alpha_f\overline{\bm{u}}_f\,\overline{\bm{u}}_f)\right)&=&-\alpha_f \nabla \overline{p} + \alpha_f\nabla \cdot\left(\mu_f\left(\nabla\overline{\bm{u}}_f+\nabla\overline{\bm{u}}_f^T-\frac{2}{3}(\nabla\cdot\overline{\bm{u}}_f)\bm{I}\right)+\bm{R}_{\mu,f}\right)\nonumber\\
      	&-&\iint_{\bm{y}\in S_I}\bm{n} \cdot(\bm{\tau}-\overline{\bm{\tau}})g(\bm{x}-\bm{y})dS-\nabla \cdot (\alpha_f\bm{\tau}_\mathrm{sfs,f}). \label{eq:appendix_3}
\end{eqnarray}

%In multiphase flow simulations where the filter width is much larger than the characteristic interface scale, the interface stresses must also be modeled. For example, in the case of a flow laden with small spheres with diameters $d_p\ll \delta_f$, velocities $\bm{u}_p$, and located at $\bm{x}_p$, the term in (\ref{eq:appendix_1}) can be approximated using drag correlations as follows:
%\begin{eqnarray}
%  \iint_{\bm{y}\in S_I}\bm{n} \cdot\bm{\tau}'g(\bm{x}-\bm{y}) &=& \sum_p\left(\iint_{\bm{y}\in S_p}\bm{n} \cdot\bm{\tau}'dSg(\bm{x}-\bm{y})\right)\\
%  &\simeq & \sum_p\left(\iint_{\bm{y}\in S_p}\bm{n} \cdot\bm{\tau}'dS\right)g(\bm{x}-\bm{x}_p)\\
%  &\simeq & \frac{1}{2}C_d \rho_f |\overline{\bm{u}}_f-\bm{u}_p|\left(\overline{\bm{u}}_f-\bm{u}_p\right)g(\bm{x}-\bm{x}_p).
%  \end{eqnarray}
%  This traditional modeling approach requires the filter width to be much larger than the characteristic interface length scale (e.g. particle diameter). While we adopt the opposite limit  $\delta_f\gg \delta_c$ in the VFIB method, the connection noted in this appendix shows how closures can be extracted by post-filtering highly-resolved VFIB simulations using coarse filters.
  }

\bibliography{references/references_himanshu,references/references_houssem}
\end{document}